\documentclass[a4paper,11pt]{article}
\pdfoutput=1 

\usepackage{jinstpub}
\usepackage{orcidlink}
\usepackage{lineno}
\usepackage{siunitx}
\usepackage[version=4]{mhchem}
\usepackage{subcaption}
\DeclareSIUnit{\photoelectron}{p.e.}
\DeclareSIUnit{\e}{e}

\title{
    Cosmic Ray Measurements Using Charge and Light Readout in a Pixelated Liquid Argon Time Projection Chamber
}

\collaboration{\includegraphics[height=17mm]{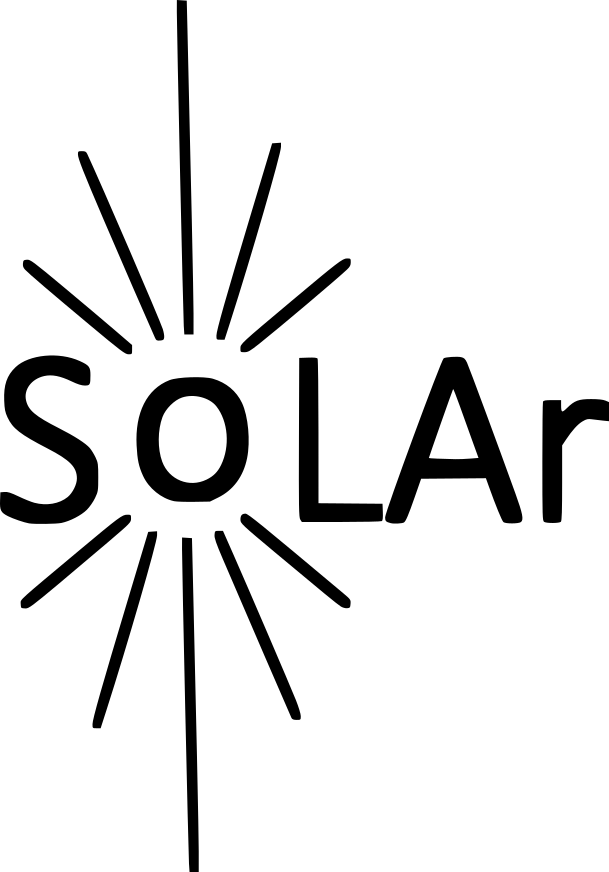}\\[6pt]
The SoLAr Collaboration}

\def\visitor{1}
\def\uniMiB{2}
\def\infnMiB{3}
\def\LHEP{4} 
\def\CIEMAT{5}
\def\imperial{6}
\def\LBNL{7}
\def\UoM{8}
\def\sheffield{9}
\def\ucBerkeley{10}
\def\QMUL{11}
\def\MIT{12}
\def\Edin{13}

\author[\visitor]{N.~Anfimov,}

\author[\uniMiB,\infnMiB]{A.~Branca,}
\author[\LHEP]{J.~B\"{u,}rgi,} 

\author[\LHEP]{L.~Calivers,} 
\author[\infnMiB]{P.~Carniti,} 
\author[\CIEMAT]{E.~Calvo,} 

\author[\infnMiB]{E.~Cristaldo,} 
\author[\CIEMAT]{C.~Cuesta,}

\author[\uniMiB, \infnMiB]{F.~Declich,}
\author[\LHEP]{R.~Diurba,} 
\author[\imperial]{P.~Dunne,} 
\author[\LBNL]{D.~A.~Dwyer,} 

\author[\UoM]{J.~Evans,} 
\author[\sheffield]{A.~C.~Ezeribe,} 

\author[\LHEP]{A.~Gauch,} 
\author[\CIEMAT]{I.~Gil-Botella,} 
\author[\infnMiB]{C.~Gotti,} 
\author[\LBNL, \ucBerkeley]{S.~Greenberg,} 
\author[\uniMiB, \infnMiB, *]{D.~Guffanti\note[*]{Corresponding author.},}  

\author[\LBNL]{A.~Karcher,}
\author[\LHEP]{J.~Kunzmann,} 

\author[\UoM]{N.~Lane,} 

\author[\CIEMAT]{S.~Manthey Corchado,} 
\author[\QMUL]{N.~McConkey,} 
\author[\uniMiB,\infnMiB]{A.~Minotti,} 

\author[\imperial]{A.~Navrer-Agasson,}  

\author[\LHEP]{S.~Parsa,} 
\author[\uniMiB, \infnMiB]{G.~Pessina,}  

\author[\UoM]{G.~Ruiz Ferreira,} 
\author[\MIT]{B.~Russell,} 

\author[\imperial]{S.~S\"{o,}ldner-Rembold,} 
\author[\Edin]{A.~M.~Szelc,} 

\author[\imperial]{A.~Tapper,} 
\author[\uniMiB, \infnMiB]{F.~Terranova,} 
\author[\LHEP]{C.~Tognina,} 
\author[\uniMiB,\infnMiB]{D.~Trotta,} 
\author[\LHEP]{S.~Tufanli,}

\author[\infnMiB]{H.~Vieira de Souza,} 
\author[\Edin]{G.~Vitti Stenico,} 
\author[\CIEMAT]{A.~Verdugo,} 

\author[\LHEP]{M.~Weber,} 

\author[\imperial]{and I.~Xiotidis} 

\affiliation[\visitor]{Affiliated with an institute covered by a cooperation agreement with CERN}
\affiliation[\uniMiB]{Universit{\`a} degli Studi di Milano-Bicocca, I-20126 Milano, Italy}
\affiliation[\infnMiB]{Istituto Nazionale di Fisica Nucleare Sezione di Milano Bicocca, I-20126 Milano, Italy}
\affiliation[\LHEP]{University of Bern, CH-3012 Bern, Switzerland}
\affiliation[\CIEMAT]{CIEMAT, Centro de Investigaciones Energ{\'e}ticas, Medioambientales y Tecnol{\'o}gicas, E-28040 Madrid, Spain}
\affiliation[\imperial]{Imperial College of Science, Technology and Medicine, London SW7 2BZ, United Kingdom}
\affiliation[\LBNL]{Lawrence Berkeley National Laboratory, Berkeley, CA 94720, USA}
\affiliation[\UoM]{University of Manchester, Manchester M13 9PL, United Kingdom}
\affiliation[\sheffield]{University of Sheffield, Sheffield S3 7RH, United Kingdom}
\affiliation[\ucBerkeley]{University of California Berkeley, Berkeley, CA 94720, USA}
\affiliation[\QMUL]{Queen Mary University of London, London E1 4NS, United Kingdom}
\affiliation[\MIT]{Massachusetts Institute of Technology, Cambridge, MA 02139, USA}
\affiliation[\Edin]{University of Edinburgh, Edinburgh EH8 9YL, United Kingdom}

\emailAdd{daniele.guffanti@cern.ch}

\abstract{Liquid argon time projection chambers have emerged as a competitive technology for detecting solar neutrinos. The SoLAr collaboration was formed to explore argon detectors with pixelated light and charge readout, aiming for high detection efficiency and improved energy resolution. Building on the success of an initial prototype, we present results obtained with a second SoLAr prototype (V2), a $30\times30\times\SI{30}{cm^3}$ time projection chamber operated in a cryostat containing several hundred kilograms of liquid argon. We report measurements of cosmic-ray muons using both tracking and calorimetry from light and charge sensors, and we highlight the improved performance achieved through combined charge and light reconstruction. These results demonstrate the promise of dual-readout detectors and motivate future prototyping efforts toward kiloton-scale facilities.}


\begin{document}
\maketitle
\flushbottom
\section{Introduction}

Nearly sixty years after their first observation, solar neutrinos remain a subject of interest for both the particle-physics and solar-physics communities.
Measurements of solar neutrino fluxes from the various nuclear reactions that power the Sun are crucial for testing solar models, which serve as benchmarks for our understanding of stellar evolution.
Among the expected reactions, only one has eluded detection so far: the fusion of a proton with \ce{^{3}He}, a process with an extremely small cross section that nevertheless produces the highest-energy neutrinos expected from the Sun, with energies reaching up to \SI{20}{MeV}.
Furthermore, a high-precision measurement of the neutrino flux produced by \ce{^{8}B} decay would make it possible to determine the solar mixing parameters, thereby testing the current \(1\text{--}2\,\sigma\) tension in the value of \(\Delta m^2_{12}\) obtained from reactor experiments, as recently highlighted by global analyses of solar neutrino data within the three-neutrino mixing framework~\cite{Capozzi:2018dat,Meighen-Berger:2024xbx,JUNO:2025gmd}.

The Deep Underground Neutrino Experiment (DUNE) intends to explore the field of solar neutrinos~\cite{Capozzi:2018dat,DUNE:2020ypp} with its kilo-tonne-scale argon-based Far Detector modules hosted at the Sanford Underground Research Facility~\cite{DUNE:2020txw,DUNE:2023nqi}, approximately \SI{1.5}{km} underground.

Three challenges exist for next-generation solar neutrino studies utilizing argon detectors.
First, radiological backgrounds must be mitigated with a combination of passive shielding and event discrimination techniques.
Second, the detector must have precise energy resolutions for neutrinos at energies on the order of 10 MeV. Initial studies have found that an adequate resolution for discoveries with solar neutrinos is approximately 7\%~\cite{Capozzi:2018dat}.
Finally, the data rates for kiloton-scale argon detectors must be reduced. Current technologies require several petabytes per year of data for MeV-scale neutrino detection, due to the high backgrounds at these energies~\cite{DUNE:2020lwj,Garcia-Peris:2025exn}. 
These requirements demand the addition of online or nearline triggering to reduce the amount of data stored. Many of these challenges are being explored by a variety of R\&D programs focusing on different technologies and target materials. DUNE has summarized many of these technologies as options for the ``Module of Opportunity" within their Phase II program~\cite{DUNE:2024wvj}.

The SoLAr detector concept is one of the possibilities being explored. It proposes a liquid argon time projection chamber (LArTPC) with a novel dual-readout anode equipped with charge- and light-sensitive sensors on the same printed circuit board (PCB). Pixelated anode tiles overcomes the possible reconstruction ambiguities of conventional 2D projective readout \cite{Adams:2019uqx}. Integrating light-sensitive detectors on the anode will enhance the light collection and its uniformity, a key requirement for exploiting a combined calorimetry approach using both ionization and scintillation signals \cite{LArIAT:2019gdz, Shi:2025rob, Ning:2024zxg}.
A first prototype of such a ``hybrid'' tile was successfully operated in 2022 confirming the feasibility of a dual-readout anode for a LArTPC~\cite{SoLAr:2024fwt}. 

This paper describes the operation and the performance of the second prototype of the SoLAr anode design, which we refer to as SoLAr V2. It features a sensitive area five larger than the first prototype, and uses the latest Hamamatsu VUV-sensitive SiPMs and charge-sensitive pixels with \SI{4}{mm} pitch instrumented by LArPix-v2b chips. SoLAr V2 took cosmic-ray data in 2023 over the timespan of a week. Thanks to its larger size and longer data-taking period, we completed a suite of calibrations using the light and charge information, including calorimetric measurements of cosmic muon events. 
These results characterize the capabilities of a dual-readout liquid argon time projection chamber with metrics inspired by previous liquid argon detectors~\cite{MicroBooNE:2019efx,DUNE:2020cqd,ICARUS:2024hmk}.

\section{SoLAr V2 Detector}
\label{sec:setup}
This section describes the design of the anode PCBs of the SoLAr V2 TPC, the charge and light read-out, and the cryogenics facility at the University of Bern. 

\subsection{Design of the Anode PCB}
\begin{figure}[htbp]
    \centering
    \begin{subfigure}[b]{0.3\textwidth}
        \includegraphics[height=4.5cm]{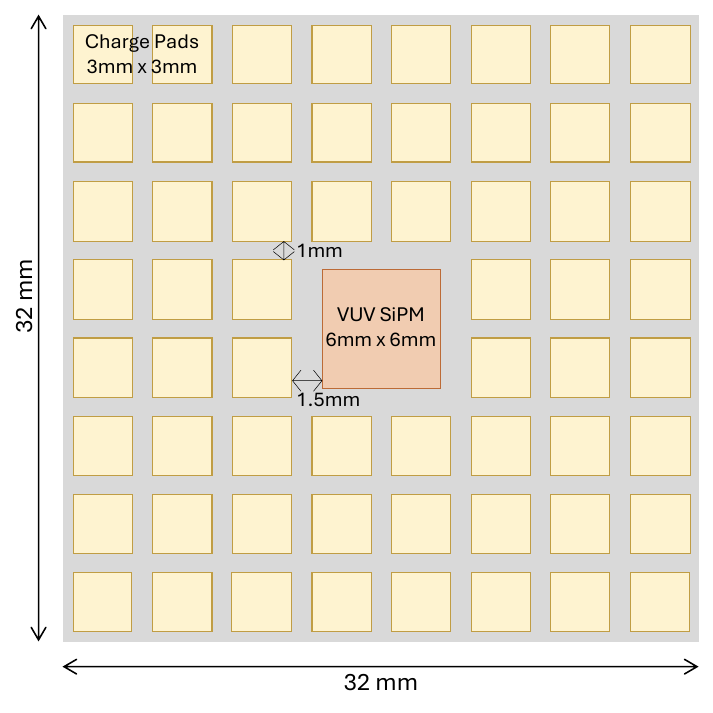}
        \caption{}
        \label{fig:solar:v2:tile:zoom}
    \end{subfigure}
    \hfill
    \begin{subfigure}[b]{0.3\textwidth}
        \centering
        \includegraphics[height=4.2cm]{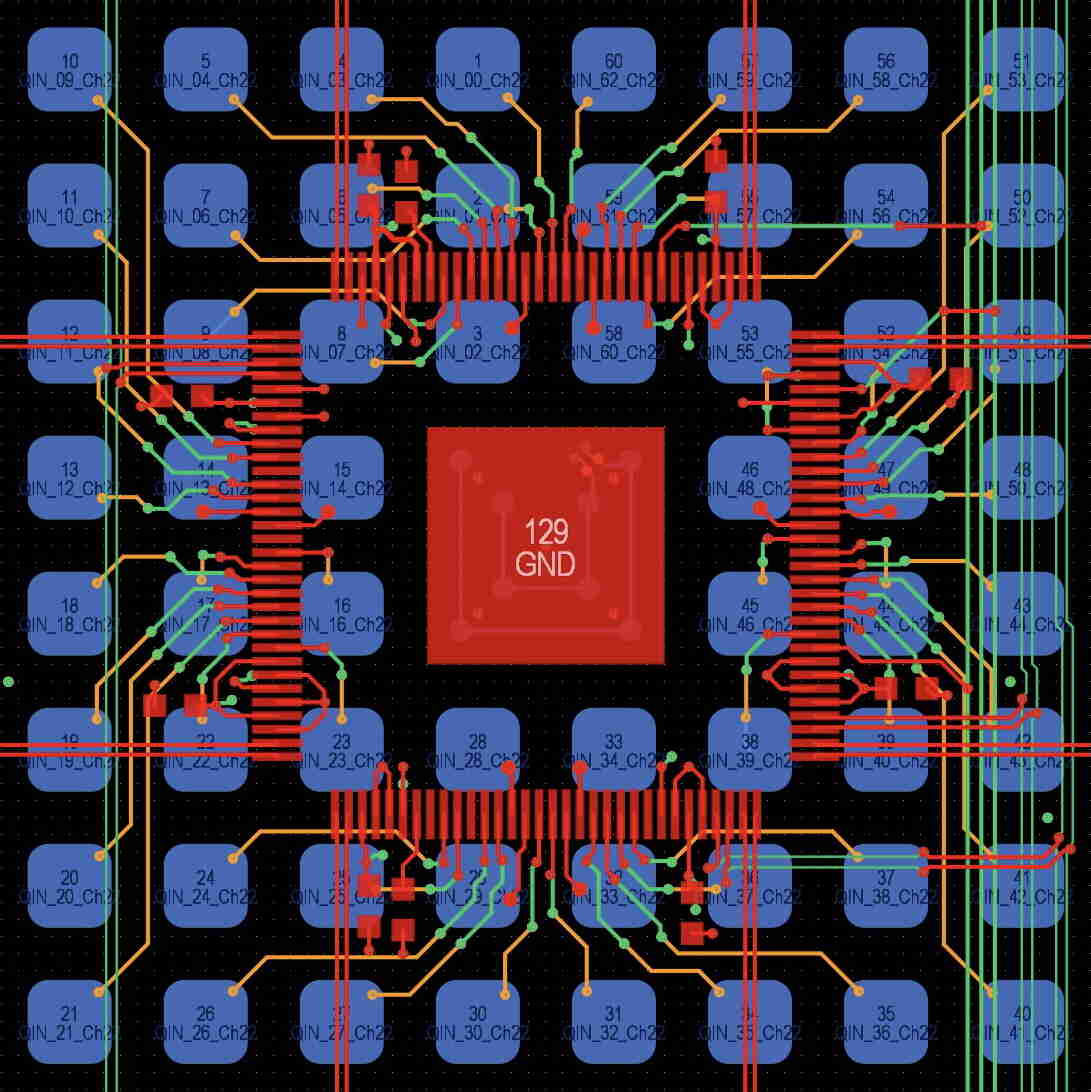}
        \vspace*{2mm}
        \caption{}
        \label{fig:solar:v2:tile:schematic}
    \end{subfigure}
    \hfill
    \begin{subfigure}[b]{0.3\textwidth}
        \centering
        \includegraphics[height=4.2cm]{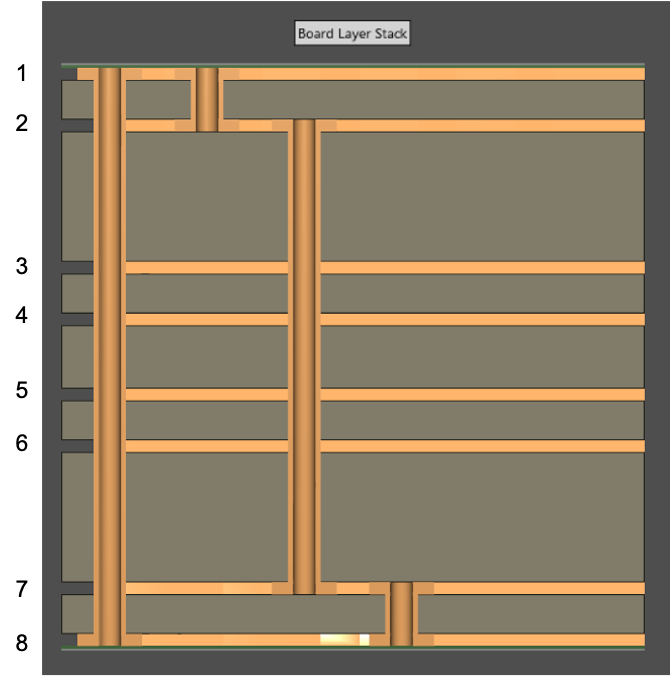}\vspace*{2mm}
        \caption{}
        \label{fig:solar_pcb_layer_stack}
    \end{subfigure}
    \caption{SoLAr v2 prototype unit cell design. (a) Each unit cell consists of one SiPM (6 mm by 6 mm) and sixty charge collection pads (3 mm by 3 mm). (b) Detailed view of the PCB traces connecting each pixel to the LArPix chip, located on the back side of the PCB. (c) Stack-up of the layers of the PCB.}
    \label{fig:solar:v2:tile}
\end{figure}

\begin{figure}[htbp]
    \centering
    \includegraphics[width=0.49\linewidth]{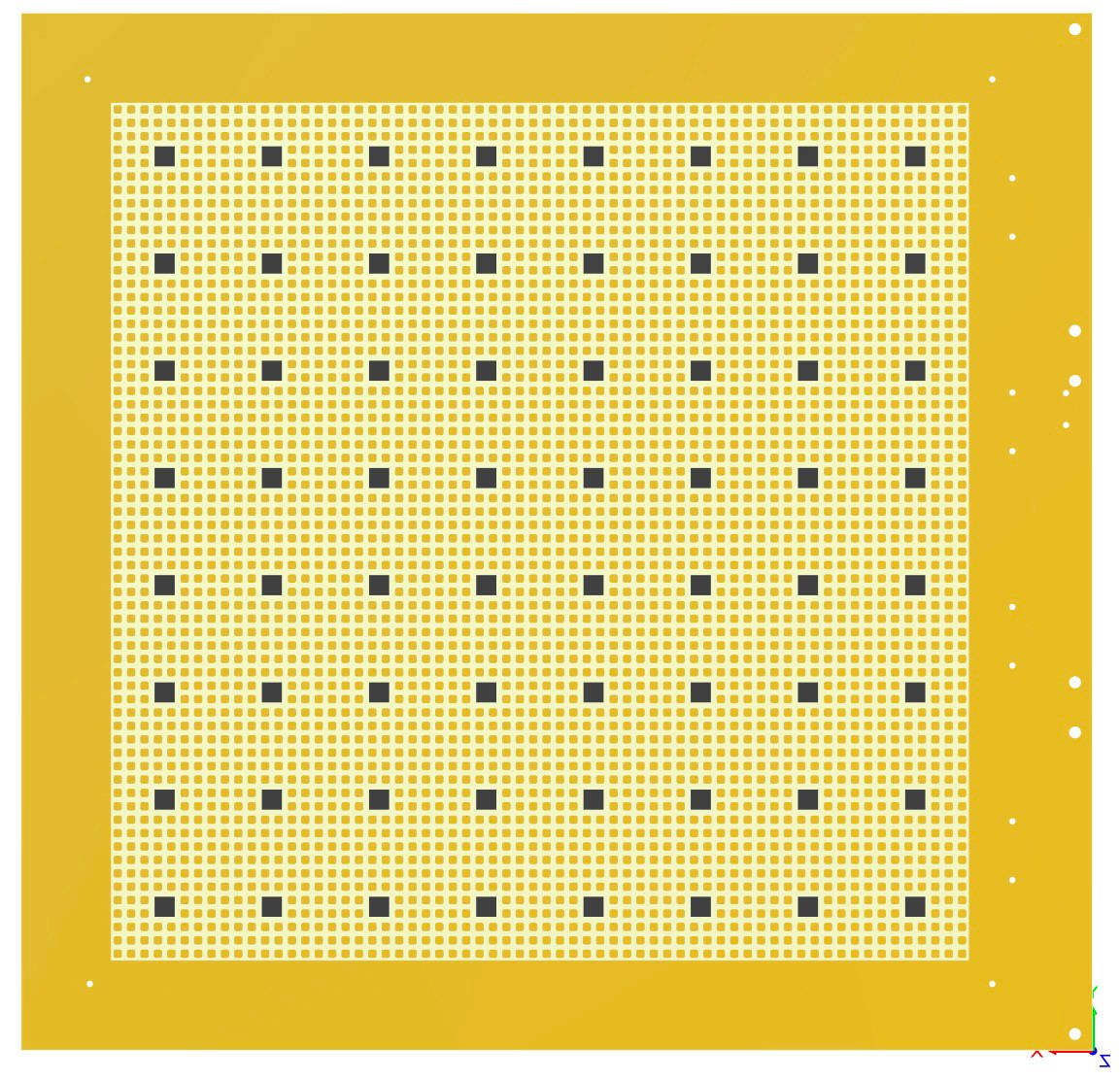}
    \hfill
    \includegraphics[width=0.48\linewidth]{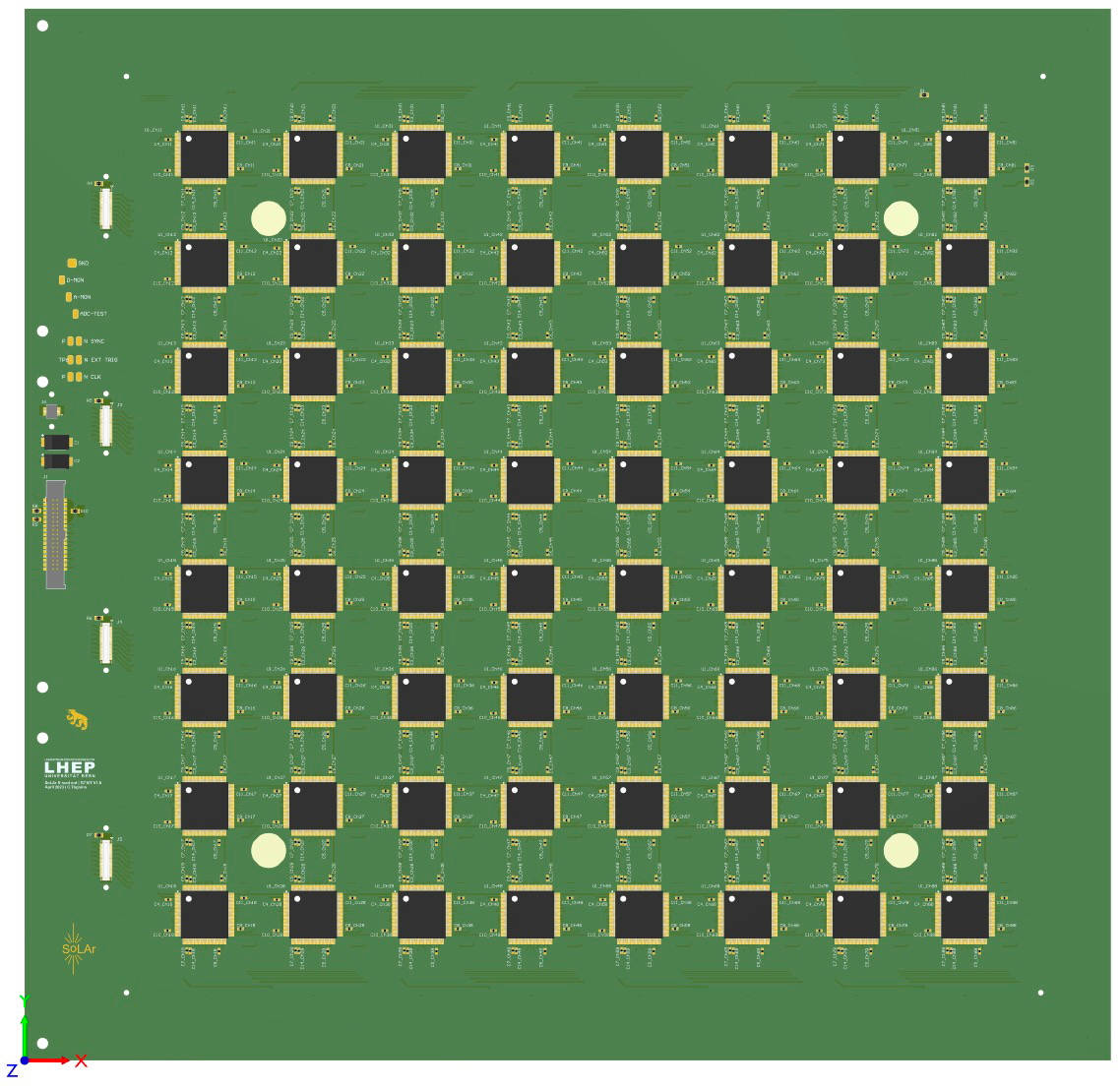}
    \caption{CAD drawing of the SoLAr V2 anode PCB.}
    \label{fig:solar:v2:PCB}
\end{figure}

The SoLAr V2 TPC has dimensions of $32.8 \times \SI{33.3}{cm^2}$ in the $x$ and $y$ directions, with a sensitive anode area of $25.6 \times \SI{25.6}{cm^2}$, and a drift distance of \SI{30.2}{cm} along the $z$ axis. The anode plane layout design is organised into a grid of 64 identical regions, referred to as SoLAr unit cells, each consisting of a single SiPM centred within a three-pixel-wide perimeter of 60 charge pixels as shown in Figure~\ref{fig:solar:v2:tile}. The charge collection pads have a dimension of $3\times\SI{3}{mm^2}$, with a pitch of \SI{4}{mm}.
The VUV SiPMs are chip-sized packaged, with through-silicon vias. They have an area of $6 \times \SI{6}{mm^2}$ and are arranged on a square grid with a \SI{32}{mm} pitch. The mounted SiPM surfaces are not aligned with the charge collection pads and protrude approximately \SI{1}{mm} from the anode plane. This design can accommodate up to 3,840 charge pads and 64 SiPMs across the entire anode plane as shown in Figure~\ref{fig:solar:v2:PCB}. The SiPMs constitute $6\%$ of the active anode plane.

The anode plane PCB, sometimes referred to as the anode tile, has a complex multilayer design to address signal routing of the charge pads to the readout ASICs mounted on the back side of the anode tile and the SiPMs to the four light readout connectors, while foreseeing sufficient shielding layers to minimize the potential crosstalk between the two readout systems. Figure~\ref{fig:solar_pcb_layer_stack} shows the PCB eight-layer stack-up. The upper layer (Layer 1) hosts the LArPix IO traces and the connectors, Layers 2 and 7 host routing of the charge traces, Layers 3 and 6 are shielding ground layers, Layer 4 hosts routing of the light signal traces and Layer 5 hosts digital and analogue voltage lines of LArPix, and Layer 8 hosts the charge collection pads.

\subsection{Charge Readout}
\begin{figure}[htbp]
    \centering
    \begin{subfigure}[b]{0.55\linewidth}
        \centering
        \includegraphics[height=3.5cm]{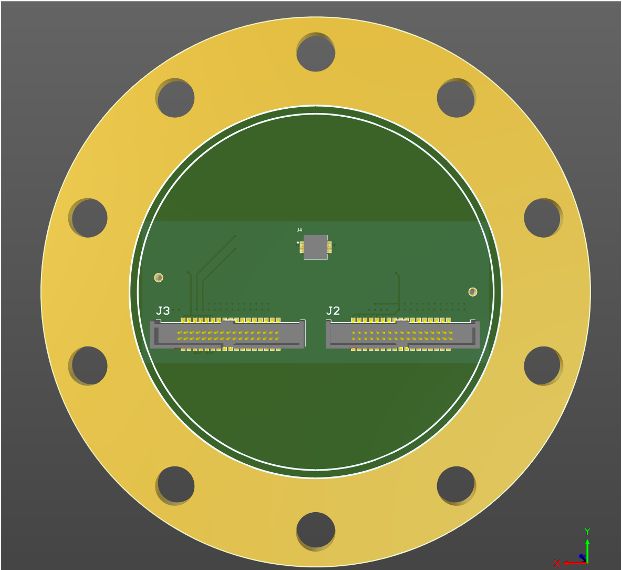} 
        \includegraphics[height=3.5cm]{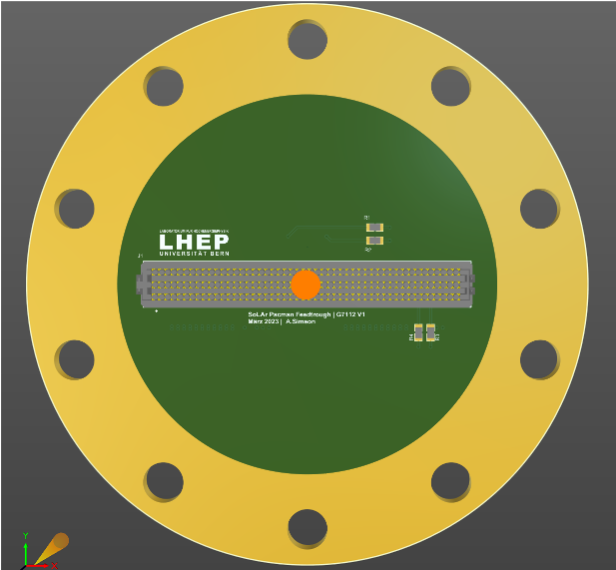}
        \caption{}
        \label{fig:feedthrough_charge_warm}
        \includegraphics[width=8cm]{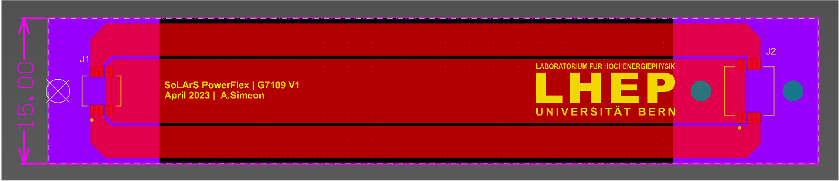}
        \caption{}
        \label{fig:flex_power_charge}
        \includegraphics[width=8cm]{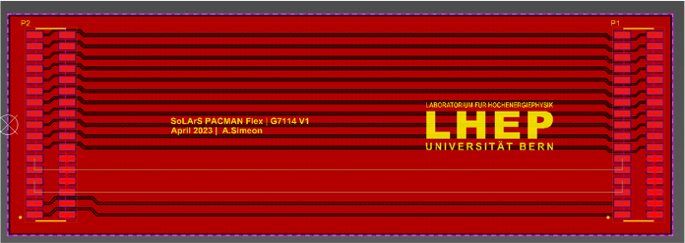}
        \caption{}
        \label{flex_data_charge}
    \end{subfigure}
    \hfill
    \begin{subfigure}[b]{0.40\linewidth}
        \centering
        \includegraphics[height=9cm]{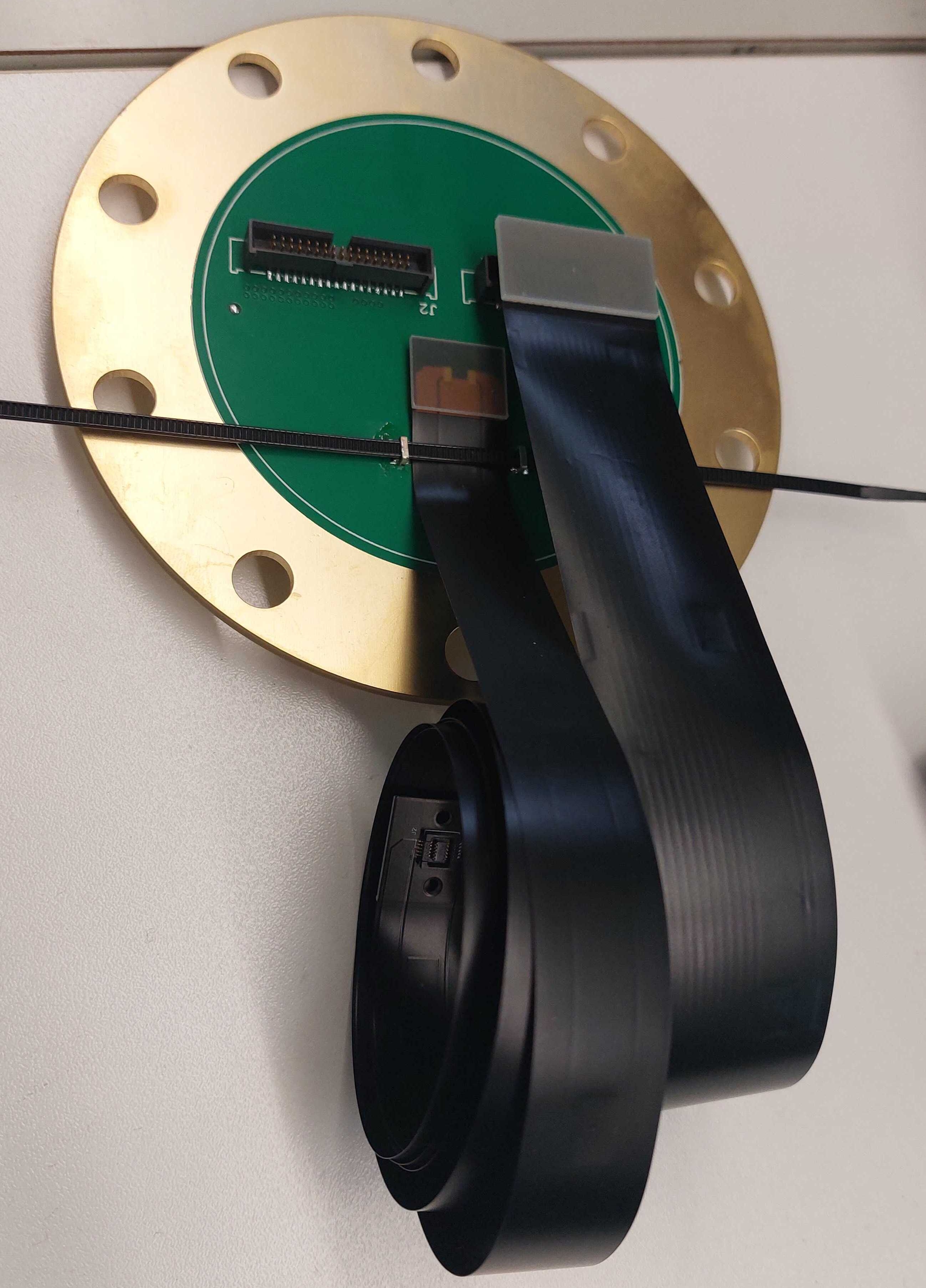}
        \caption{}
        \label{fig:feedthrough_charge}
    \end{subfigure}

    \caption{(a) Charge feed-through, (b) FPC cable for charge power length: 1.2 m, width 1.5 cm, (c) FPC cable for charge data length: 1.2 m, width: 2.5 cm (d) The charge feed-through and cables.} 
    \label{fig:charge_feedthrough}
\end{figure}

The charge readout system is instrumented with LArPix-v2b chips~\cite{Dwyer:2018phu}. 
The LArPix chip provides a scalable cryo compatible readout with low power analogue front-end of $< \SI{200}{\micro W}$/channel and highly multiplexed digital I/O with up to 10240 channels/cable.
Each LArPix ASIC features 64 low noise, self triggering channels with tunable thresholds. 
The LArPix chips are connected across the anode tile through programmable Hydra networks and are linked to the warm controller via four root chips. Two custom-designed 1.2 m long flexible printed circuit (FPC) cables are used for charge data and power transmission from the anode tile to the charge feed-through, as shown in Figure~\ref{fig:charge_feedthrough}. 
A warm controller called PACMAN is mounted on the warm side of the charge feedthrough and provides charge system clock, power, and digital I/O for control and detector data. The data is streamed continuously over ethernet to the host machine.

Due to a shortage of supplies, only 20 of the 64 available slots on the SoLAr V2 anode tile could be instrumented with LArPix chips. Therefore, only 1,200 of the possible 3,840 pixels are read out on the anode plane with an active surface for charge-sensitive pixels of $16.0\times \SI{12.8}{cm^2}$. The non-instrumented charge pads are all connected using copper tape and grounded to avoid accumulation of drift electrons on their surface during operation. Figure~\ref{fig:solar:v2:tpc:open} shows the 20 cells with instrumented pixels. 

\subsection{Light Readout}
\begin{figure}[htbp]
     \centering
     \hspace{1cm}
     \begin{subfigure}[b]{0.40\linewidth}
     \centering
         \includegraphics[height=8cm]{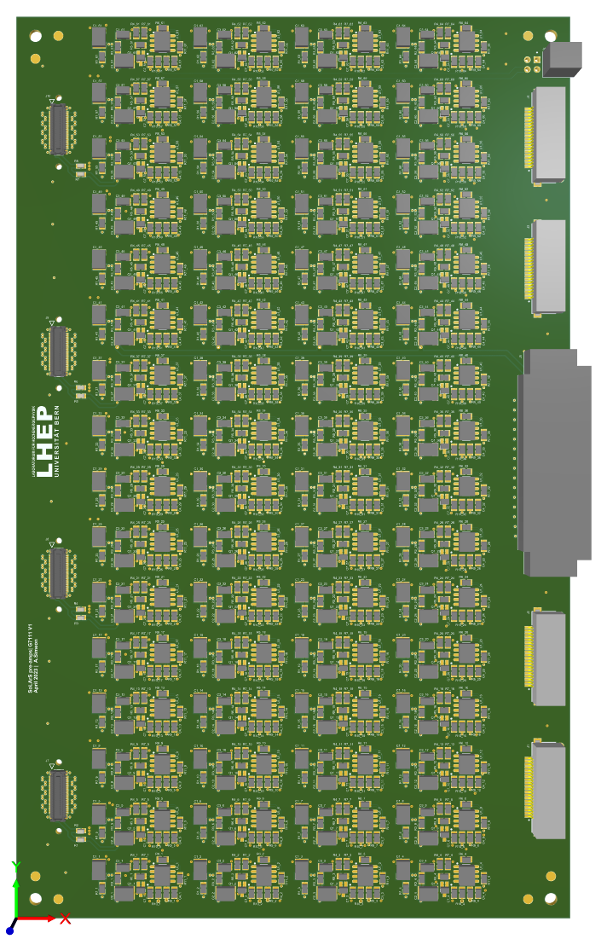}
         \caption{}
         \label{fig:solar:v2:preamp}
     \end{subfigure}
     \hfill
     \begin{subfigure}[b]{0.4\linewidth}
     \centering
         \includegraphics[height=8cm]{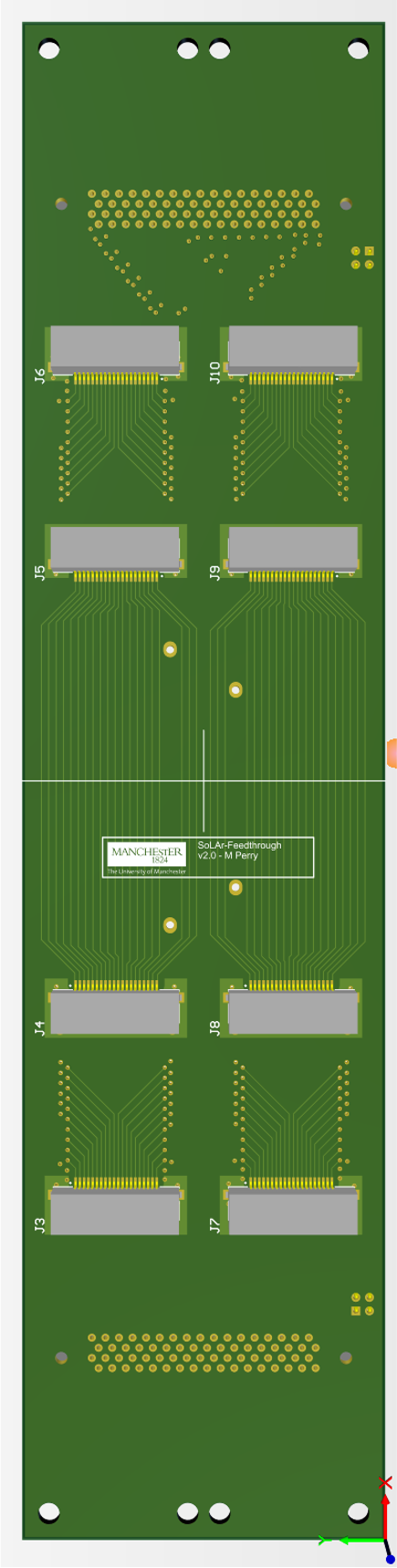}
         \caption{}
         \label{fig:solar:v2:light:feedthrough}
     \end{subfigure}
     \begin{subfigure}[b]{\linewidth}
         \centering
         \includegraphics[width =\linewidth]{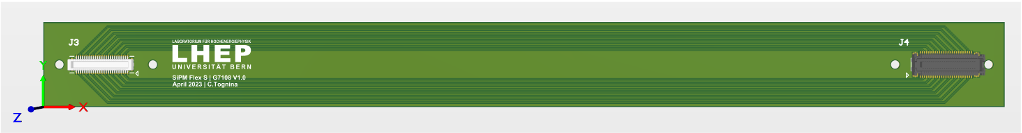}
         \caption{}
         \label{fig:flex_data_light}
     \end{subfigure}
     \caption{(a) The cold preamplifier board; (b) light feed-through; (c) FPC cable for light data with a length of \SI{18}{cm} and a width of \SI{1.6}{cm}.}
 \end{figure}

For the light readout of the SoLAr V2 prototype, we use the latest Hamamatsu VUV SiPMs, type S13370-6050CN, which is part of Hamamatsu's VUV4 family of SiPMs~\cite{SiPMdatasheet}. These sensors are chip-sized, edgeless, surface-mount devices manufactured using through-silicon-via technology. The photon detection efficiency (PDE) of the SiPMs for liquid-argon scintillation light is quoted as approximately $15\%$ at cryogenic temperature for an overvoltage of \SI{4}{V} and has been verified in multiple experimental studies \cite{Pershing:2022eka, Alvarez-Garrote:2024byb}.
The 64 SiPMs on SoLAr V2 are divided into groups of 16. The traces of 16-SiPM groups are routed to a dedicated light-readout connector on the anode-tile PCB. A diagram of the connector is shown in Figure~\ref{fig:solar:v2:light:feedthrough}.

Four custom-designed \SI{18}{cm} FPC cables, highlighted in Figure~\ref{fig:flex_data_light}, connect the anode tile to the cold preamplifier board, which is firmly attached to the back of the anode-support G10 plate, as shown in Figure~\ref{fig:solar:v2:tpc:back}. Each cable provides the SiPM bias voltage and transmits the signals from 16 SiPMs.

The FPC cable connectors are secured on both ends by screws that fasten them to the corresponding PCBs. The preamplifier circuit, based on an LMH6624 chip, shapes and amplifies the analog SiPM signals by a factor of 10 when powered with $\pm\SI{5}{V}$. The signals are then transmitted via four SAMTEC data cables to the light feedthrough. The total power consumption of the preamplifier board is approximately \SI{7.4}{W}. The light feedthrough, shown in Figure~\ref{fig:solar:v2:light:feedthrough}, hosts connectors for the preamplifier and SiPM bias voltages, as well as four SAMTEC data connectors on both the warm and cold sides. On the warm side, the signals are routed to a variable-gain amplifier (VGA) tunable in the range $0$--–$24$~\si{dB}, before being digitized and read out by a \SI{62.5}{MSample\per s}, 14-bit analog-to-digital converter (ADC) that samples 1000 values of each SiPMs waveform, resulting in a \SI{16}{\micro s} time window.

\subsection{TPC Assembly and Integration}
\begin{figure}[htbp]
\centering
\begin{subfigure}[]{0.26\textwidth}
    \centering
    \includegraphics[height=6cm]{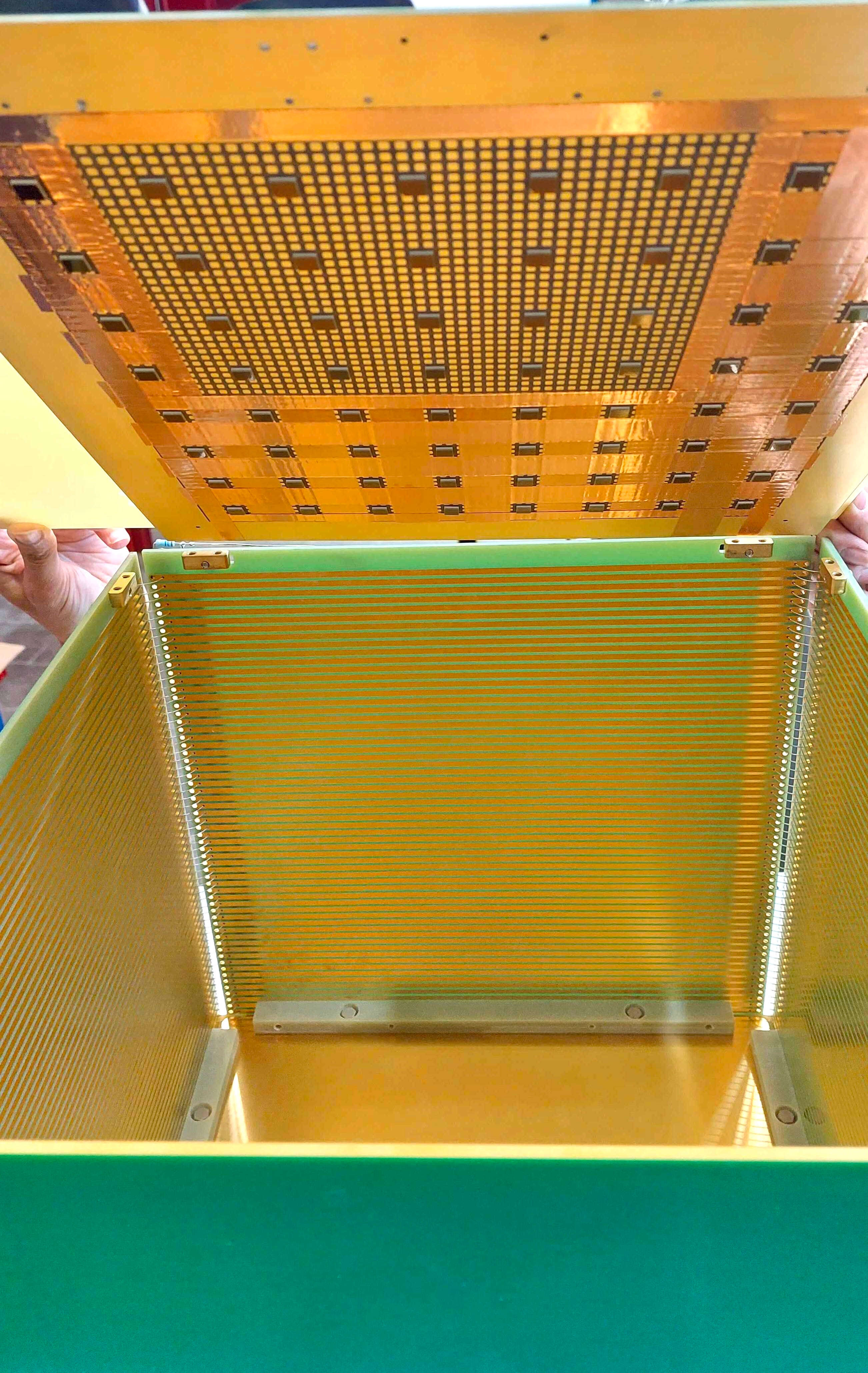}
    \caption{}
    \label{fig:solar:v2:tpc:open}
\end{subfigure}
\begin{subfigure}[]{0.4\textwidth}
    \centering
    \includegraphics[height=6cm]{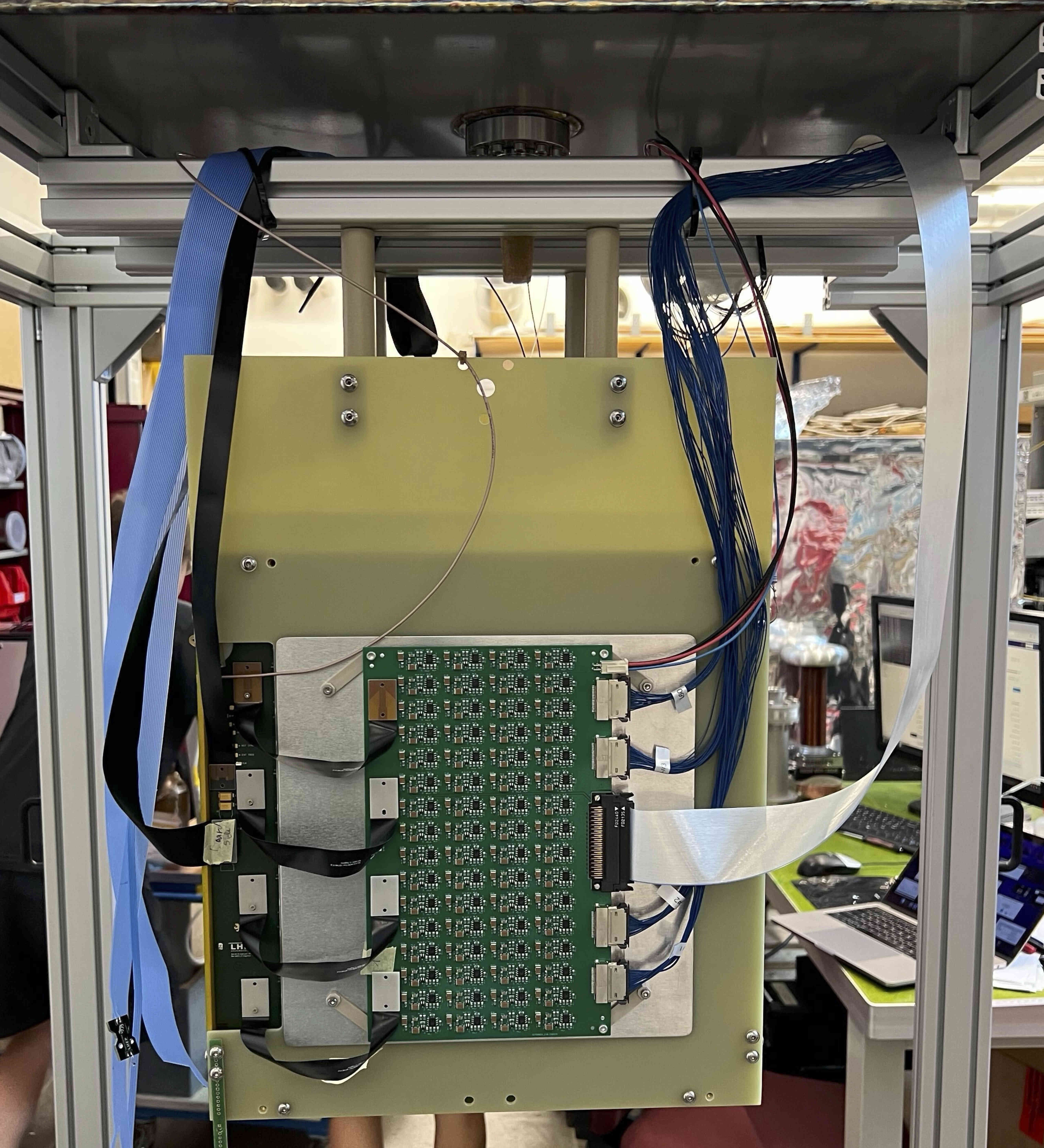}
    \caption{}
    \label{fig:solar:v2:tpc:back}
\end{subfigure}
\begin{subfigure}[]{0.26\textwidth}
    \centering
    \includegraphics[height=6cm]{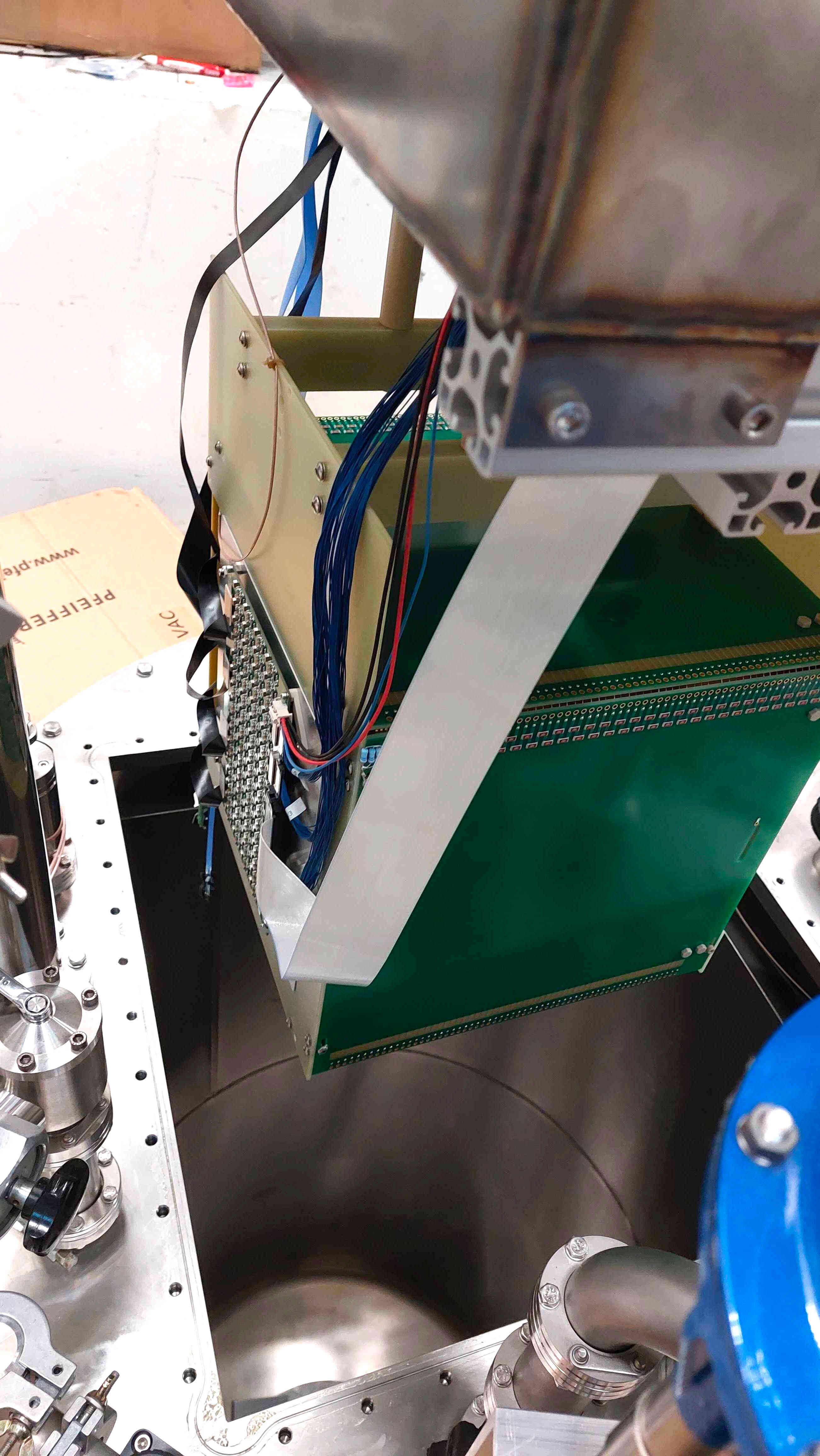}
    \caption{}
    \label{fig:solar:v2:tpc:closed}
\end{subfigure}

\caption{(a) Interior view of the TPC, illustrating the arrangement of the anode plane, cathode, and field cage; (b) the fully assembled and cabled TPC with the cold preamplifier PCB mounted and visible on the support structure; (c) SoLAr V2 TPC before insertion into the cryostat.}
\label{fig:solar:v2:tpc}
\end{figure}

Figure~\ref{fig:solar:v2:tpc:open} provides a view of the interior of the TPC, showing the anode plane, cathode, and field cage, while Figure~\ref{fig:solar:v2:tpc:closed} shows the fully assembled TPC, suspended from the top flange of the cryostat, ready for insertion. The SoLAr V2 TPC was designed to operate in the same cryostat used as a part of the DUNE Near Detector prototyping program~\cite{DUNE:2024fjn}, and is therefore equipped with a liquid-argon purification system to ensure stable long-term operation.

\subsection{DAQ and Operations}
\label{sec:solar:DAQ}

The SoLAr detector concept relies on the accurate detection and processing of light and charge signals to capture the data necessary for event reconstruction. The readout electronics are designed to manage these signals efficiently, ensuring synchronisation and precise timing for effective correlation during offline analysis.

\begin{figure}
    \centering
    \includegraphics[width=\linewidth]{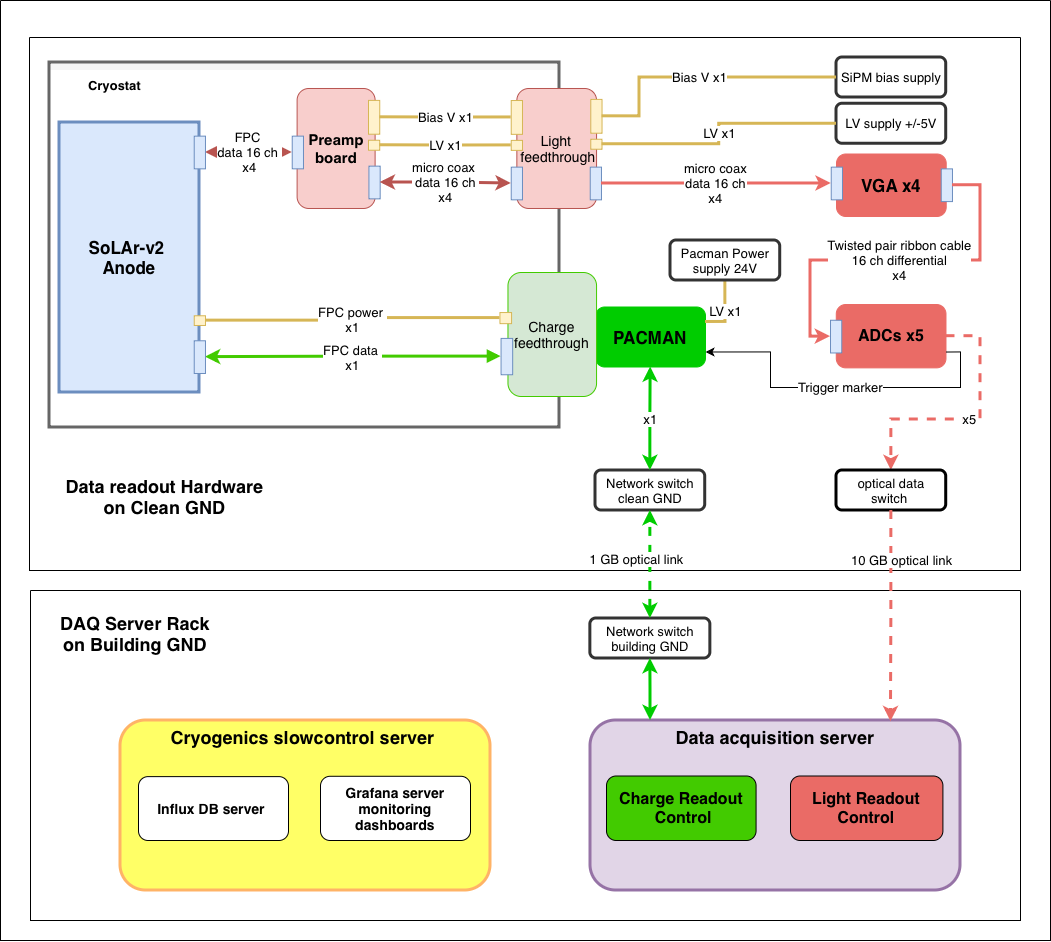}
    \caption{Diagram of the SoLAr V2 data acquisition system.}
    \label{fig:solar:readout}
\end{figure}

The SoLAr system continuously collects charge and light data through the self-triggering data acquisition (DAQ) system, which is described in Figure~\ref{fig:solar:readout}. To synchronise charge and light readout, a pulse-per-second signal from the Global Positioning System (GPS) is provided to both systems. A light trigger signal is also sent to the charge readout PACMAN, inserting a timestamp into the charge data stream to mark the start of each event. The timestamp from the light information ensures the correct alignment of the charge drift times. Both charge and light signals are time-stamped with universal time stamps, enabling their integration into coherent ``events'' during offline processing. The system is identical to the DAQ used in the DUNE Near Detector prototyping operations outlined in Ref.~\cite{DUNE:2024fjn}.

The cryogenics run of SoLAr V2 took place in the week of 3--10 July 2023. After two days for cooldown, filling, and system turn on, the data taking campaign was carried out with nominal HV of \SI{15}{kV}, corresponding to \SI{0.5}{kV/cm} drift electric field. At the end of the cosmic-ray run, a \ce{^{60}Co} source was installed outside the cryostat to study the performance of the detector at the MeV energy scale. The detector was exposed to the source, which produces two $\gamma$ rays of 1.17~MeV and 1.33~MeV in cascade, for a few hours.

\section{Detector Calibration}
\label{sec:calibration}
\subsection{Charge readout system}
\label{sec:cali:charge}

The charge readout system is calibrated by scanning each channel through periodic diagnostic runs. These tests are adapted from the DUNE Near Detector prototypes, which also use the LArPix readout system~\cite{DUNE:2024fjn}. During these diagnostic runs, triggers are sent to sample each channel on a single ASIC, and the returned value from the analog-to-digital converter (ADC) is forwarded to the data acquisition system. In calibration mode, the ASICs internally generate forced triggers.

The results are used to determine the pedestal value on a channel-by-channel basis, setting the ADC baseline and correcting for channel-to-channel variations. The overall gain function used to convert ADC values to ionized electrons is \SI{245}{e^{-}\per mV}, as measured in bench tests for the DUNE Near Detector prototypes~\cite{DUNE:2024fjn}.

\begin{figure}[htb]
    \centering
    \begin{subfigure}[b]{.48\linewidth} 
        \includegraphics[width=\textwidth]{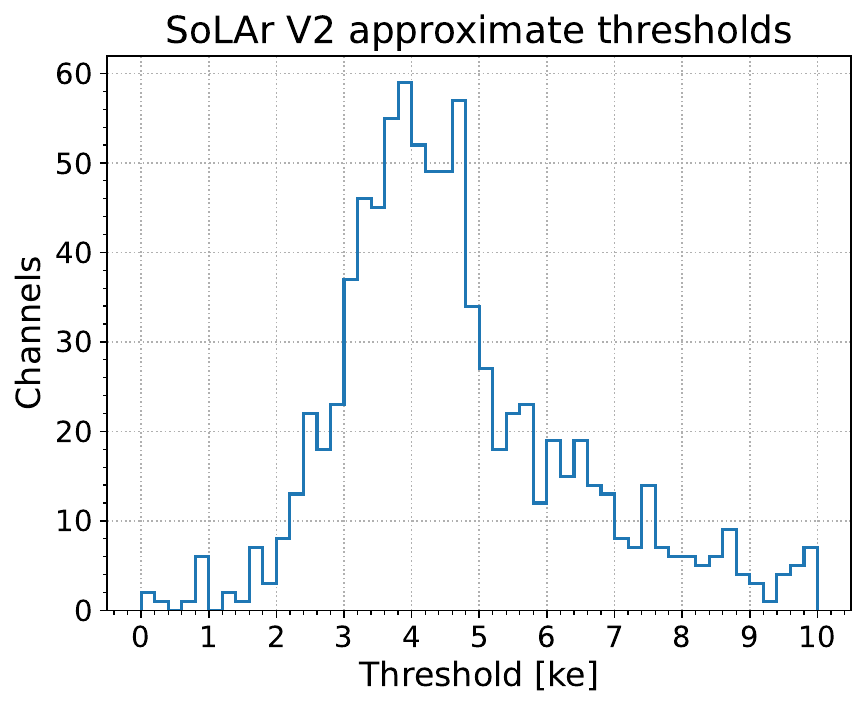}
        \caption{}
        \label{fig:avgKE}
    \end{subfigure}
    \hfill 
    \begin{subfigure}[b]{.48\linewidth}
        \includegraphics[width=\textwidth]{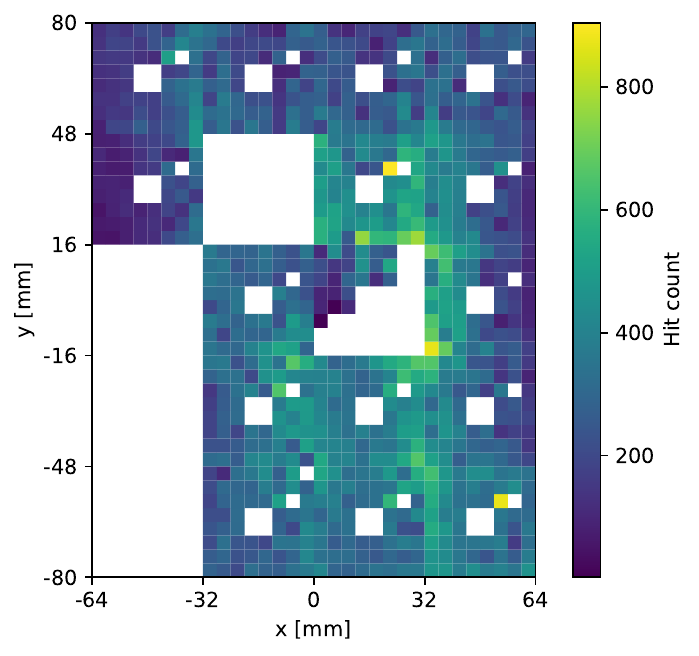}
        \caption{}
        \label{fig:activeChanelMap}
    \end{subfigure}
    \hfill
    \caption{%
        (a) Distribution of thresholds for the charge-sensitive pixels expressed in terms of thousand of electrons collected on the chip as estimated from SoLAr V2 diagnostic runs.
        (b) Number of charge hits collected across the charge-sensitive pixels,
        highlighting active and inactive areas.}
\end{figure}

Within the same diagnostic tests, we also evaluate the per-channel thresholds by disabling the periodic reset of the front-end, allowing the leakage current to integrate until a hit is triggered. The resulting charge, affected by the intrinsic noise of both the front-end and the internal discriminator, provides an approximate estimate of each channel’s threshold. The distribution of threshold values for all 910 active channels is shown in Figure,\ref{fig:avgKE}.

Figure~\ref{fig:activeChanelMap} shows the active channels and the corresponding signal distribution observed from a subset of cosmic-ray muon data. Some pixels were inactive during operation due to high noise, which resulted in them being masked. This feature is also employed for the DUNE Near Detector system~\cite{DUNE:2024fjn}. The missing pixels in the center of the detector and on the top-right of each SiPM were masked for this reason. With additional time, a higher threshold could have been applied to keep these pixels active without complications; however, due to time constraints, they were left masked.

An advantage of the LArPix system is its ability to reroute data transmission to avoid inactive chips, specifically the four chips on the left side of the anode plane in Figure~\ref{fig:activeChanelMap}. However, due to time limitations during data taking, we did not fully utilize this feature to reconfigure the readout network for this cryogenics run. Recent improvements to the LArPix system have enabled faster configuration and debugging of inactive chips, making it highly likely that the next iteration will have all available chips active.

\subsection{Light readout system}
\label{sec:cali:light}

The light detection system, comprising 64 SiPMs and the respective readout electronics, is characterized using dedicated calibration runs acquired during the data taking. The noise level of each channel is studied by lowering the SiPMs bias to \SI{40}{V}, well below the breakdown voltage. The recorded waveform samples do not feature any physical signal from the SiPMs and are therefore representative of the electronic noise induced by the readout chain.

\begin{figure}[hbt]
  \begin{center}
    \includegraphics[width=0.85\textwidth]{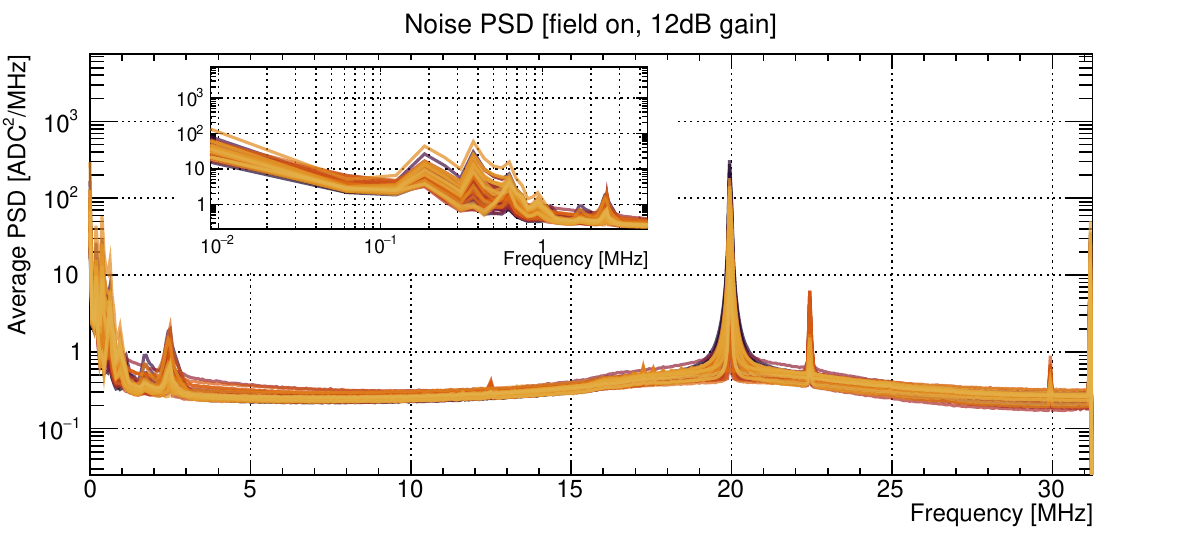}
  \end{center}
  \caption{
    Noise average power spectral density of the 64 channels composing the light detection system measured with a
    \SI{500}{V/cm} electric field applied to the TPC volume. 
    The inset offers a better view of the low-frequency region.
  }
  \label{fig:noise_psd}
\end{figure}

Figure \ref{fig:noise_psd} shows the average Power Spectral Density (PSD) of the 64 SiPM channels in presence of a \SI{500}{V\per m} electric field obtained setting a \SI{12}{dB} gain to the VGA boards. 
The readout channels exhibit a uniform behaviour in terms of noise 
for frequencies higher than \SI{4}{MHz}, featuring characteristic peaks in the 
power spectrum at \SI{20}{}, \SI{22.5}{}, and \SI{30}{MHz}. 
The difference between the noise PSD observed without the electric field if below $5\%$ for the entire frequency range. 

\begin{figure}
    \centering
    \begin{subfigure}{.49\textwidth}
        \includegraphics[width=\textwidth]{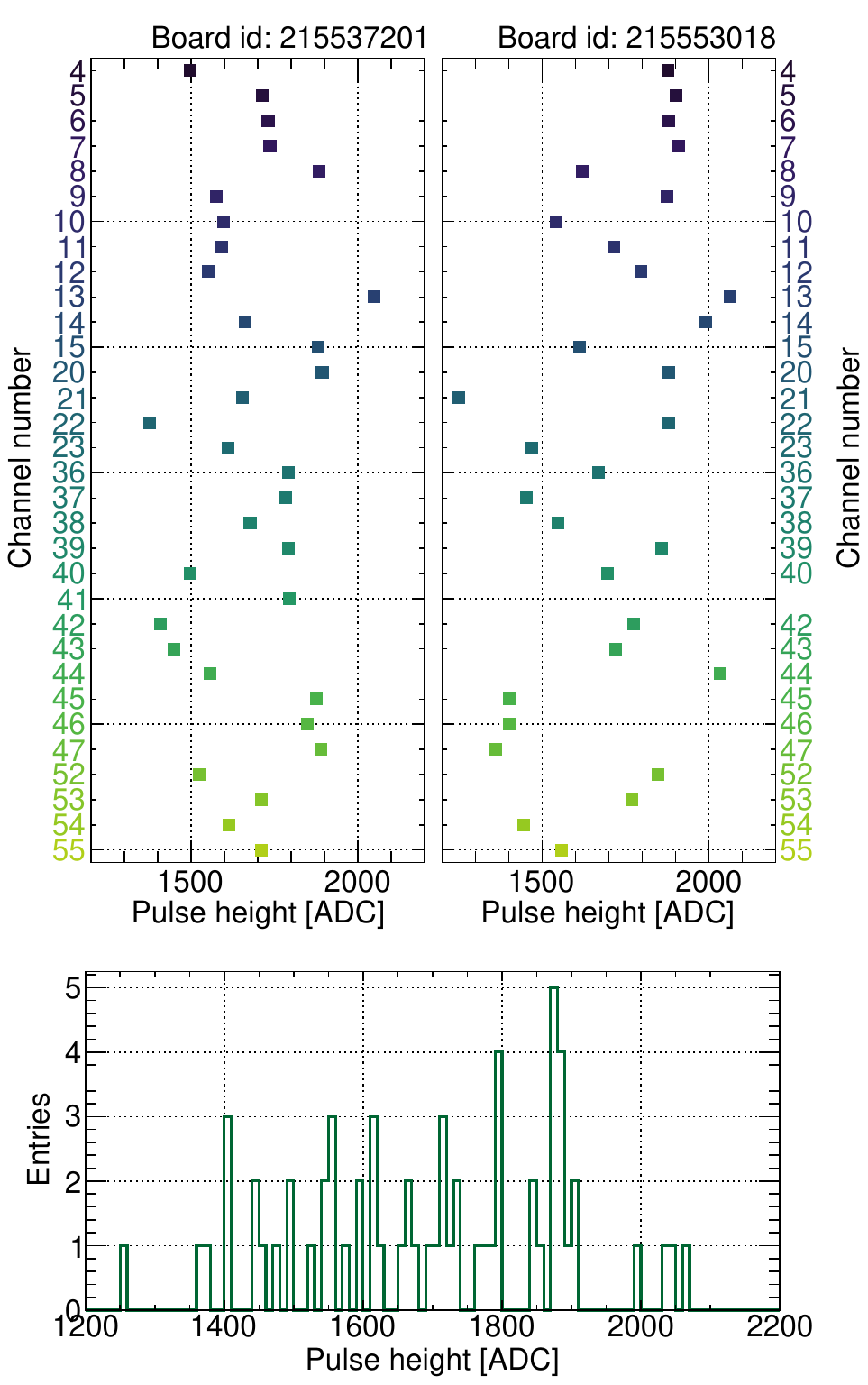}
        \caption{}
        \label{fig:ph_values}
    \end{subfigure}
    \hfill
    \begin{subfigure}{.49\textwidth}
        \includegraphics[width=\textwidth]{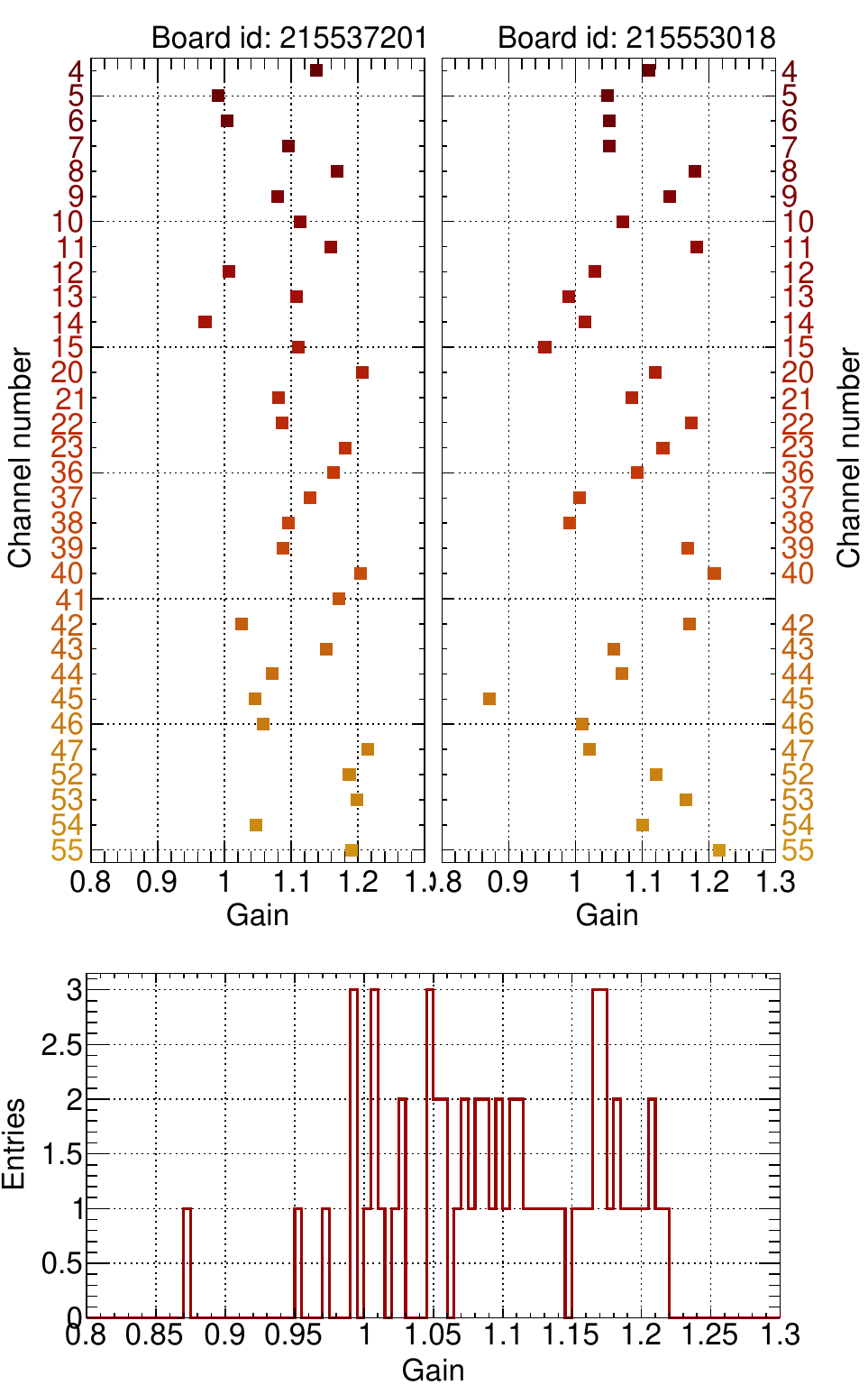}
        \caption{}
        \label{fig:gain_values}
    \end{subfigure}
    \caption{%
        (a) Top panel: values of the pulse height for single p.e.\ events for each SiPM channel as measured in the LED calibration run. Bottom panel: distribution of the pulse height values. 
        (b) Top panel: values of the gain calibration obtained for all SiPM channels. Bottom panel: distribution of the gain values.
        (a) and (b) are based on the same dataset, which has been collected biasing the SiPM at \SI{46}{V} and setting a\SI{24}{dB} gain with the VGA board. 
    }
\end{figure}

\begin{figure}
    \centering
    \begin{subfigure}{.49\textwidth}
        \includegraphics[width=\textwidth]{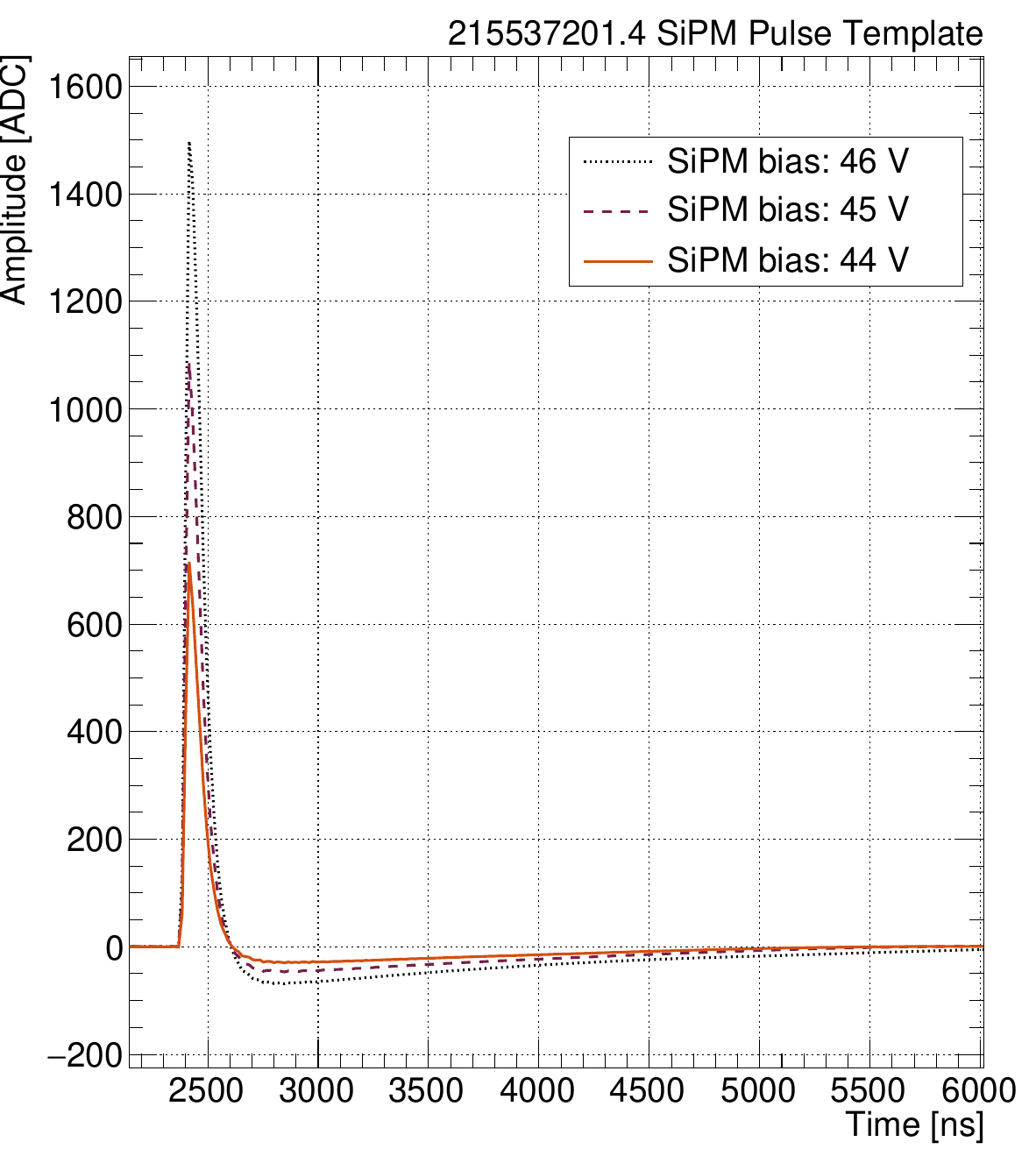}
        \caption{}
        \label{fig:sipm_template_ov}
    \end{subfigure}
    \hfill
    \begin{subfigure}{.49\textwidth}
        \includegraphics[width=\textwidth]{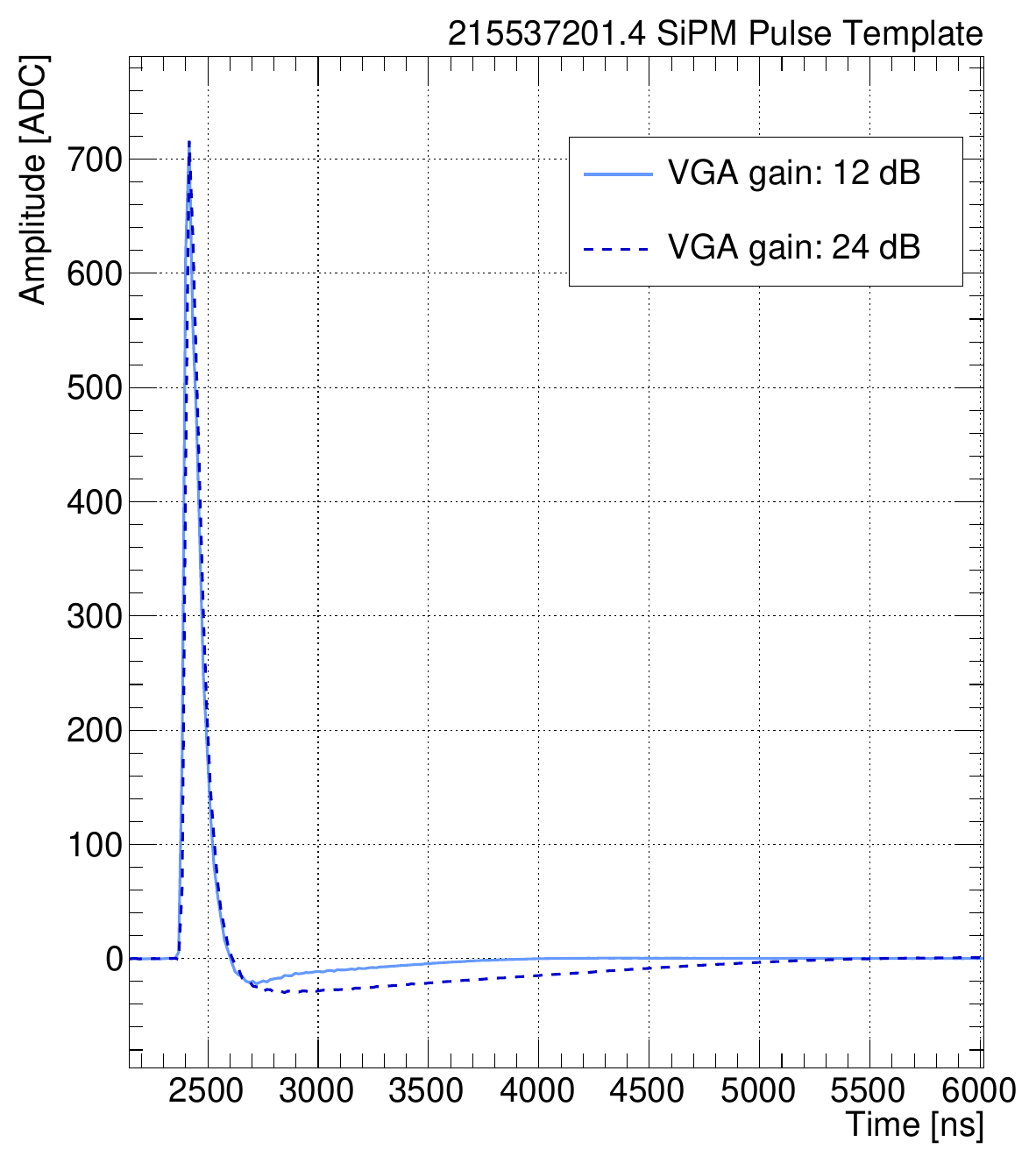}
        \caption{}
        \label{fig:sipm_template_vga}
    \end{subfigure}
    \caption{%
        Impulse response templates for a representative channel derived from multi-p.e.\ pulses and rescaled to single-p.e.\ amplitude.
       Panel (a) shows the effect of the SiPM bias voltage, while in (b) we highlight the impact of gain settings on the signal shape. In this case the template amplitudes are scaled to allow for an easier comparison.
    }
    \label{fig:sipm_template}
\end{figure}

We characterize the response of each channel to light signals by flashing the anode with an {\small LED} diode and synchronously triggering the waveform acquisition. To avoid loss of information caused by the limited depth of the 14-bit ADC, we construct the template of the impulse response of each SiPM using pulse candidates corresponding to 10 to 20 photo-electrons~(p.e.).
Using the same dataset, we build the distribution of the SiPMs pulse height by selecting the maximum amplitude of the signals in the LED flash time interval after applying a Gaussian filter. 
Figure\,\ref{fig:ph_values} reports the values of the pulse height cause by single p.e.\ events for each SiPM channel, obtained with a Gaussian fit of the corresponding peak in the pulse height distribution; these values are then used to rescale the amplitude of the SiPM response template.
The impulse response for a single channel is shown in Figure\,\ref{fig:sipm_template_ov} for a common bias voltage of 44, 45, and \SI{46}{V} applied to all SiPMs. The electronic response presents an undershoot after the pulse caused by the AC coupling of the SiPM with the cold amplifier stage.

We also characterize the impact of the VGA board bandwidth on the SiPM pulses. Figure\,\ref{fig:sipm_template_vga} shows the impulse response template for a representative channel applying a \SI{44}{V} bias and setting the VGA gain to \SI{12}{} (solid, light blue) and \SI{24}{dB} (dashed, blue), with the latter featuring a much slower baseline restoration.

\begin{figure}
    \centering
    \includegraphics[width=1.0\linewidth]{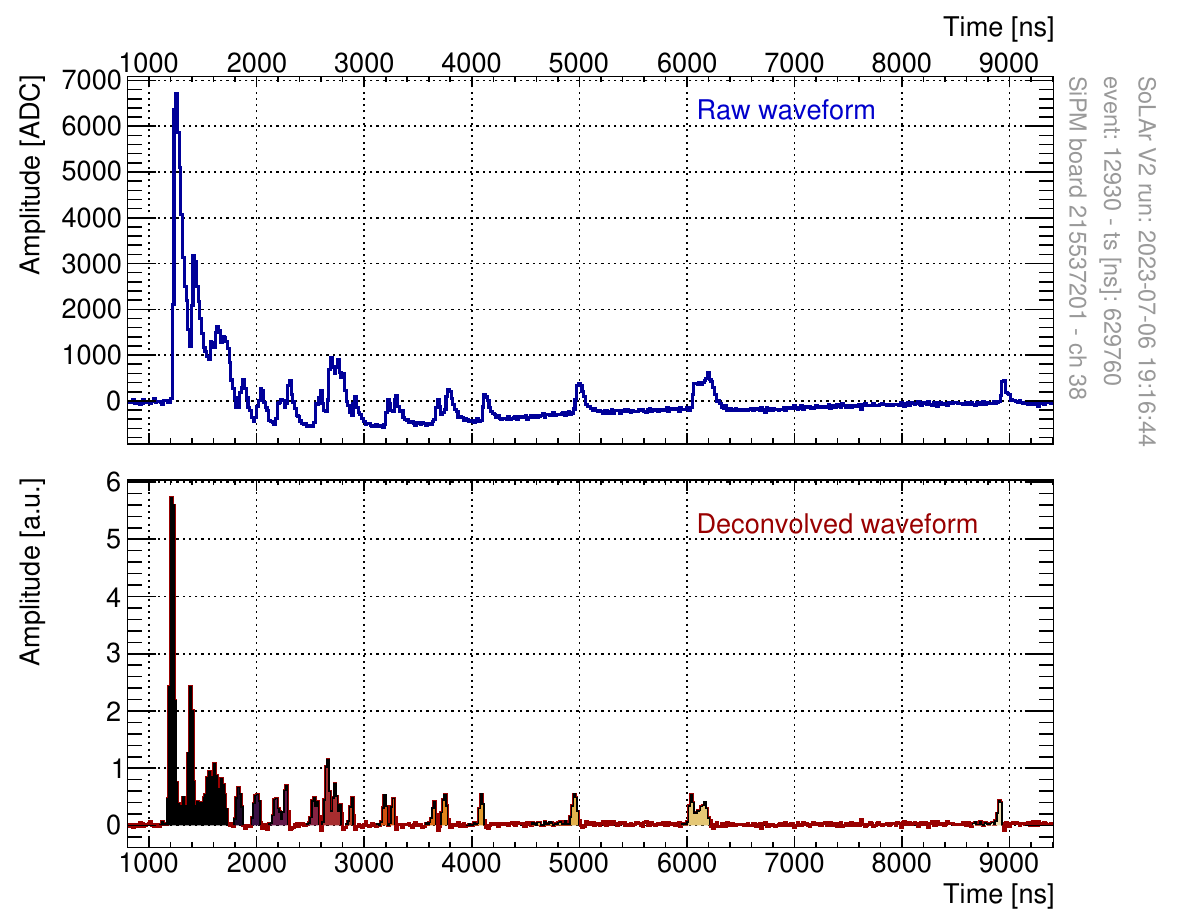}
    \caption{Effect of the deconvolution procedure applied to a raw waveform (top panel)
    acquired during the cosmic-ray data taking. 
    The deconvolved waveform (bottom panel) gives a better estimate of the original 
    signal shape and is almost unaffected by the characteristic undershoot 
    of the electronic response.
    The photon hits reconstructed in the deconvolved waveform are highlighted as a filled area.
    }
    \label{fig:filter:example}
\end{figure}

This distinctive bipolar behaviour of the impulse response causes issues in the detection of late scintillation photons emitted in the de-excitation of LAr excimers in the triple state. We correct this effect by applying an offline digital filter which performs a deconvolution of the SiPMs response, recovering with good approximation the features of the original scintillation signal. Using the response templates obtained from the {\small LED} runs, we compute for each channel $i$ a filter in the frequency domain defined as 
\begin{equation}
    W^{(i)}[f] = G[f] \cdot \frac{1}{H^{(i)}[f]} \; ,
    \label{eq:deco:sbnd}
\end{equation}
where $H^{(i)}[f]$ is the discrete Fourier Transform of the pulse response template for channel $i$ and $G[f]$ is a low-pass Gaussian filter, with its cut-off frequency set at \SI{8}{MHz}. The effect of the filter in Equation\,\eqref{eq:deco:sbnd} is demonstrated in Figure\,\ref{fig:filter:example}, which shows the digitized signal of a channel 
recorded during the cosmic-ray data taking. The deconvolution almost perfectly cancels the undershoot affecting the raw signal.

\begin{figure}[htb]
\centering
    \includegraphics[width=0.85\textwidth]{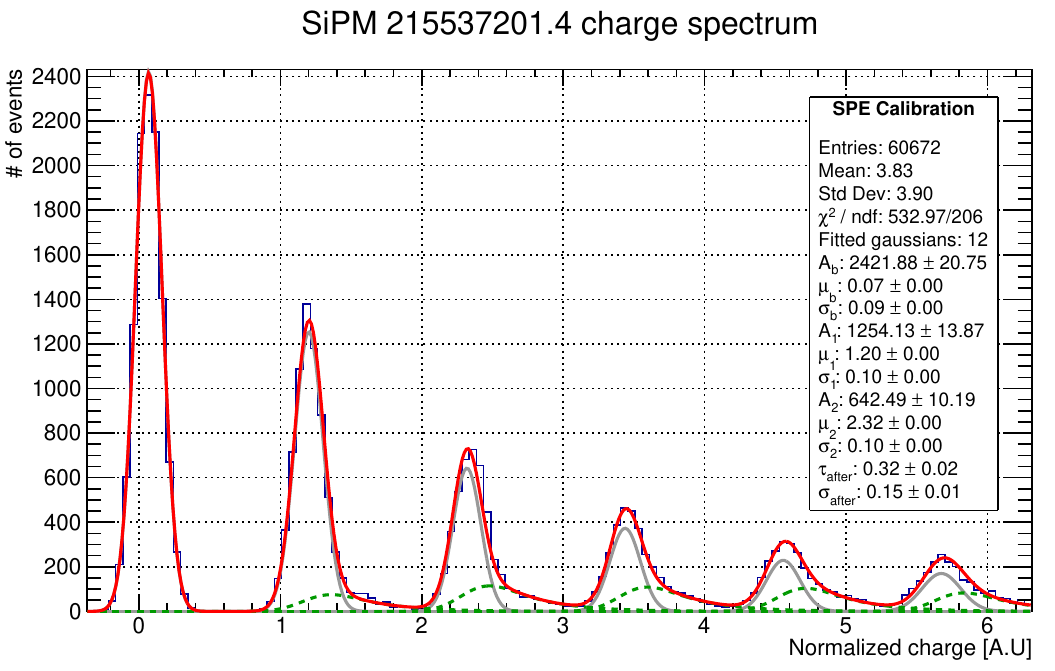}
    \caption{
        Spectrum from the {\small LED} calibration dataset for a representative SiPM biased at \SI{44}{V} ($\approx\SI{2}{V}$ above the breakdown voltage).
        In the fit model only the baseline, the 1\,p.e.\ and 2\,p.e.\ peaks are assigned three free parameters. 
        All the other Gaussian function have amplitude and standard deviation as free parameters, 
        while the gain constrains the mean. The after-pulse contributions are
        shown in green.
    } 
    \label{fig:finger_deconv}
\end{figure}

Using a model of the single p.e.\ response in the filter defined 
in Equation\,\eqref{eq:deco:sbnd} leads to pulses with an area corresponding
to the number of p.e. Nevertheless, systematic uncertainties in the construction of the single p.e.\ template and the presence of the low-pass filter require an additional 
calibration. 
The calibration is performed by measuring the gain of the SiPMs and fitting distributions of the integrated charge in the SiPM waveform after deconvolving the signals from the LED light dataset.
The fit result for a single channel is shown in Figure\,\ref{fig:finger_deconv}, where the gain is defined by the difference between the mean signal of
the 2\,p.e.\ and 1\,p.e.\ peaks. 
The presence of asymmetric tails affects the p.e.\ peaks,
likely due to after-pulsing. These effects are modelled as exponential 
functions with a Gaussian smearing added to the fit model.
The gain values obtained for all SiPM channels are shown in Figure\,\ref{fig:gain_values}

\section{Measurements of Cosmic-ray Muons}
\label{sec:analysis}
\begin{figure}[htb]
    \centering
    \includegraphics[width=\linewidth]{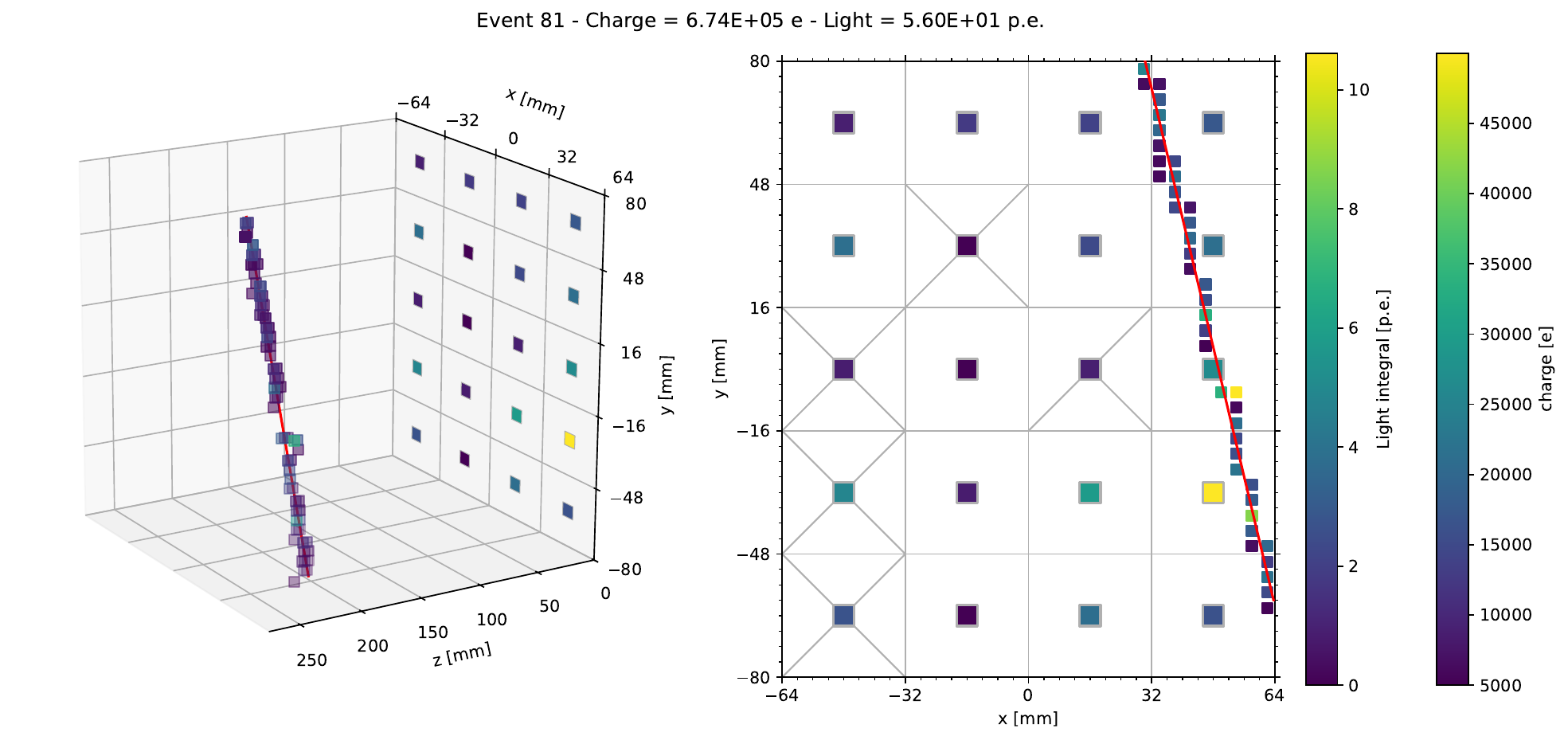}
    \caption{Event display of a cosmic muon track detected by the SoLAr V2 prototype, highlighting the correlation of light and charge information by signal intensity. Crossed sectors indicate regions of inactive charge-sensitive pixels, while SiPMs appear as central squares within each sector, with the surrounding pixels capturing ionization signals. Corrections to the location of the muon along the drift direction are applied using the time of entry of the muon from the scintillation light detected by the SiPMs at nanosecond timescales. The drift across a distance of $300$~cm takes approximately 180~$\mu$s}
    \label{fig:reco:ed}
\end{figure}

This section presents measurements of light and charge data on a selected sample of 80 minutes of cosmic muon data recorded with the SoLAr V2 prototype, containing approximately 50 thousands cosmic-ray muon tracks. Due to the small size of the detector, these GeV-scale muons typically traverse the entire TPC, depositing energy at a nearly constant rate per unit length ($\mathrm{d}E/\mathrm{d}x$) along a straight path as Minimum Ionising Particles (MIPs). Through-going muons at MIP energies provide a calibration tool through their well-understood energy deposits and straight trajectories in the detector. 
A representative event display of a muon crossing the SoLAr V2 TPC, produced using the reconstruction framework presented in this section, is shown in Figure~\ref{fig:reco:ed} to demonstrate the typical features of MIPs events in the detector.

\subsection{Performance from the Light Readout System}
\label{sec:cosmics:light}

The trigger scheme of the Light Readout System is based on a discriminator acting on the analogue sum of each 6 SiPM channels with a threshold corresponding to approximatively 2\,p.e. 
When the Light Readout System's trigger is fired, the DAQ records the 64 waveforms of the SiPMs and a TTL signal is sent to PACMAN, which will be recorded in the charge data stream as a light trigger marker to be associated with the charge signal of the same event in offline analysis.

\begin{figure}[htb]
    \centering
    \includegraphics[width=0.9\linewidth]{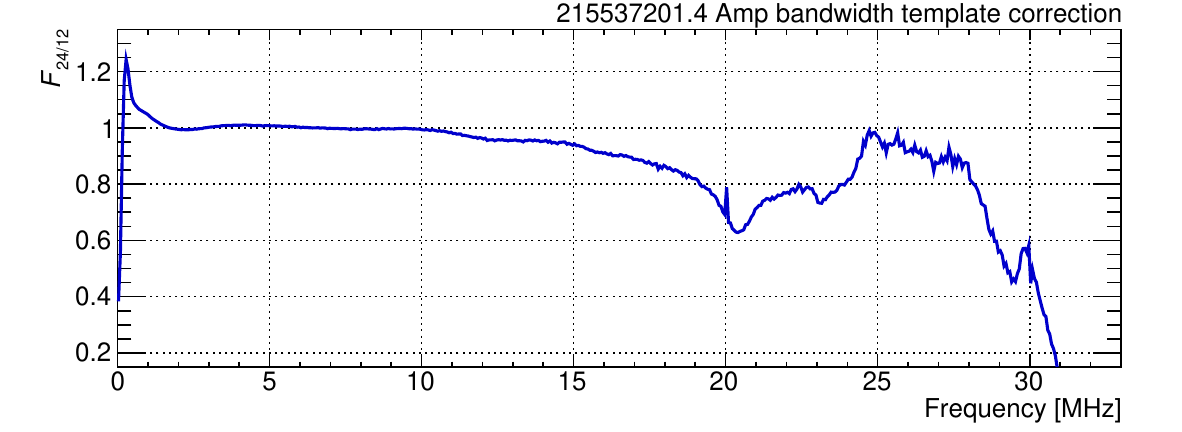}
    \caption{Correction term $F_{24/12}$ applied to the deconvolution filter for a representative channel. The correction accounts for the different bandwidth of the amplification stage in the construction of the impulse response template, which differs between the LED calibration run and the cosmic data-taking.}
    \label{fig:filter_bandwidth}
\end{figure}

During data taking, the SiPMs were biased at \SI{46}{V} (approximatively $\SI{4}{V}$ above the breakdown voltage) 
and the VGA gain was set at \SI{12}{dB}. This choice turned
out to be not optimal, as the high over-voltage worsened the 
impact of after-pulsing already visible at \SI{44}{V} in Figure\,\ref{fig:finger_deconv}. 
Furthermore, the combination of \SI{46}{V} bias and \SI{12}{dB}
gain was not studied in the {\small LED} calibration run, 
preventing us from building a model of the single-photoelectron pulse response with those settings. 
As a consequence, we use the response template obtained from the {\small LED} run at \SI{46}{V} and \SI{24}{dB} gain
in constructing the waveform filter described in Section\,\ref{sec:cali:light}, correcting the effect of the different gain setting by adding to the filter in Equation\,\eqref{eq:deco:sbnd} an additional term defined as
\begin{equation}
    F^{(i)}_{24/12}[f] = \left|\frac{H^{(i)}_\text{24\,dB}[f]}{H^{(i)}_\text{12\,dB}[f]}\right| \; ,
\end{equation}
with $H^{(i)}_{k\,\text{dB}}[f]$ being the Fourier transform of the response template obtained
from the LED run at \SI{44}{V} with \SI{12}{} and \SI{24}{dB} gain. 
An example of the $F^{(i)}_{24/12}[f]$ term as a function of the frequency 
is shown in Figure\,\ref{fig:filter_bandwidth}.

After applying the deconvolution procedure, we scan the resulting waveform looking for pulses.
Whenever a pulse is found, we define it as a photon hit and estimate its charge 
by integrating the waveform until the baseline level is restored.
Individual photon hits in the example waveform are highlighted in Fig.\,\ref{fig:filter:example}.

\begin{figure}
\centering
\begin{subfigure}{.49\textwidth}
\includegraphics[width=\textwidth]{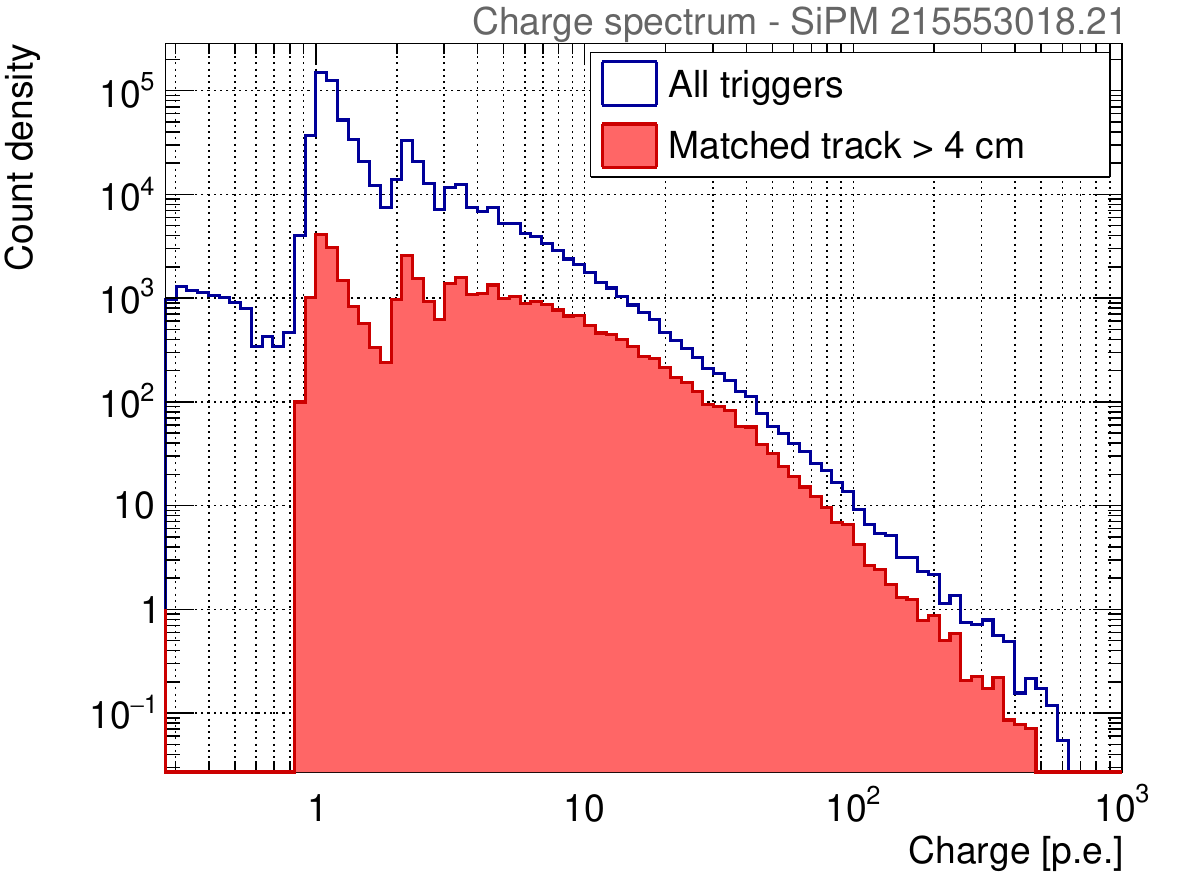}
\caption{Single SiPM charge spectrum}
\label{fig:sipm:qdataCh4}
\end{subfigure}
\hfill
\begin{subfigure}{.49\textwidth}
\includegraphics[width=\textwidth]{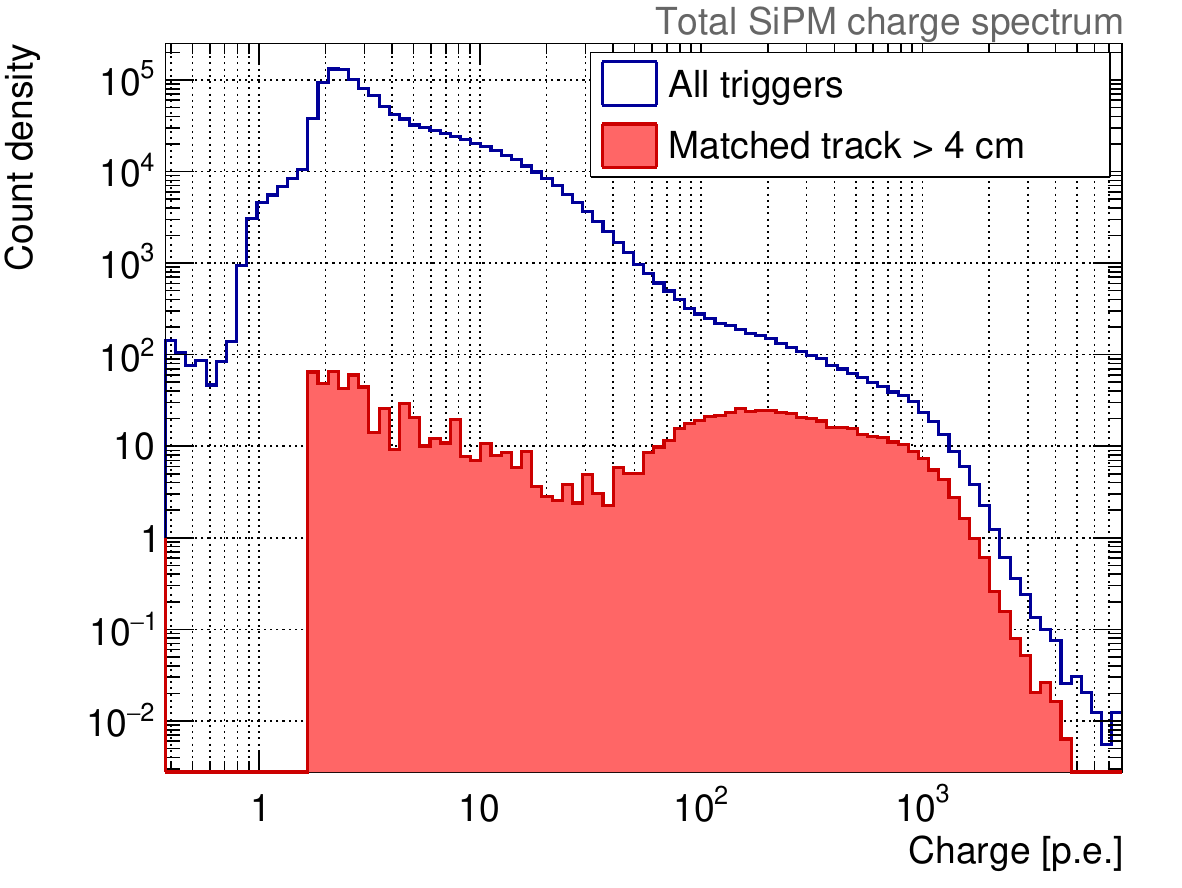}
\caption{Total SiPM charge spectrum}
\label{fig:sipm:qdataTotal}
\end{subfigure}
\caption{%
    (a) Integrated charge collected by a single SiPM and (b) by summing all SiPM channels for events recorded during the cosmic data-taking.
    The number of events in each bin is represented by its area and not by its height. 
    The red histogram shows events where we reconstruct a single cosmic-ray track candidate with the charge readout system with a track length $>4$~cm. 
}
\label{fig:sipm:qdata}
\end{figure}

Figure\,\ref{fig:sipm:qdataCh4} shows
the charge spectrum of a single channel from cosmic data taking,
where the charge has been computed by adding up all hits 
detected in a single waveform, and Figure\,\ref{fig:sipm:qdataTotal} shows
the distribution of the total charge collected by the 64 SiPMs in a single event. 
In Figure\,\ref{fig:sipm:qdataTotal} only few events are observed with a total charge below the $\approx 2$\,p.e.\
trigger threshold, caused by fake hits that are later removed in the waveform analysis procedure. 

Given the low threshold, the acquisition could be triggered both by cosmic rays crossing the TPC and by radioactive decays happening close to the anode plane. Figure\,\ref{fig:sipm:qdata} also shows the charge distribution for a single SiPM (\ref{fig:sipm:qdataCh4}) and for the entire SiPM array (\ref{fig:sipm:qdataTotal}) for those events where we reconstruct a single cosmic-ray track candidate with the charge readout system (Sec.\,\ref{sec:cosmics:charge}), with a track length $>4$~cm. 
Since only a portion of the pixelated anode area was active at the time of data-taking, this selection misses a significant fraction of cosmic rays crossing the TPC. This event sample shows that, depending on the track distance from the anode plane, the total collected charge from the SiPM array ranges from a few tens of p.e.\ to a few thousand.

\begin{figure}
    \centering
    \includegraphics[width=0.9\textwidth]{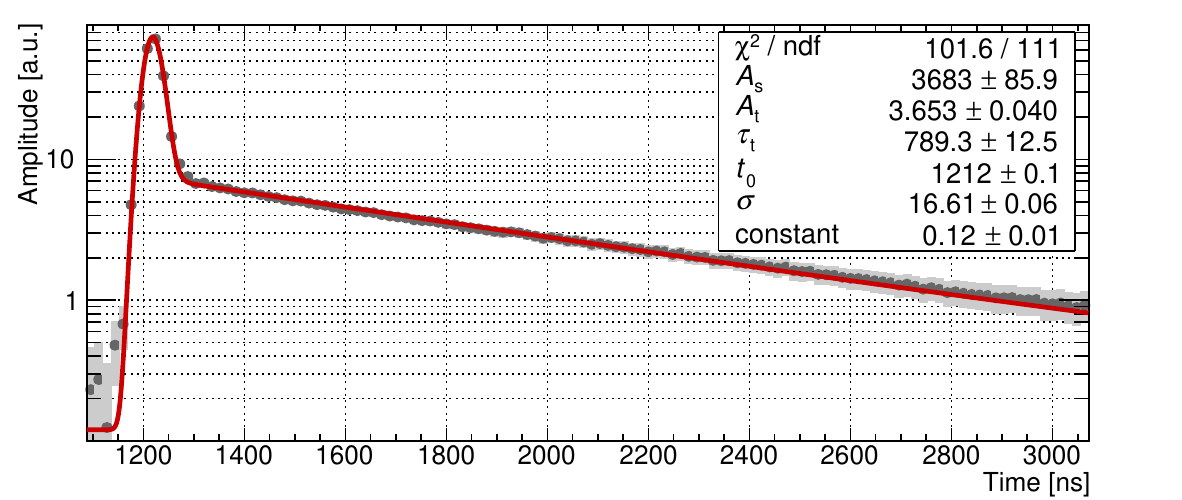}
    \caption{Liquid argon scintillation time profile built by averaging the deconvolved SiPMs' waveforms that could be match to a single muon track at least \SI{4}{cm} long reconstructed by the charge readout system. The time axis is restricted to the fit region.}
    \label{fig:avg_mu_pulse}
\end{figure}

The light signals that are matched with a single muon track are used to study the LAr scintillation time profile. For each event we construct a time profile of the scintillation flash by summing all the deconvolved waveform of the SiPMs that detected at least one hit. By averaging the flash profiles of all events detecting at least 10 p.e.\ we obtain the scintillation time evolution shown in Figure\,\ref{fig:avg_mu_pulse}, featuring the double exponential structure of LAr scintillation.

We fit the time profile with the sum of two exponential function convolved with a Gaussian to reproduce the finite time resolution of the detector. 
The time constant of the fast component from the de-excitation of the \ce{Ar_2^*} excimers in the singlet state is too short for the analysis to be sensitive to it, therefore the corresponding parameter was fixed to its nominal value of \SI{7}{ns}. 
The Gaussian approximation of the time response and the assumption of a perfectly flat baseline (i.e., perfect cancellation of the undershoot) do not capture entirely the behaviour of the system, resulting in a bad goodness-of-fit score. 
To account for unknown effects, we increased he uncertainties on each point of the time profile by a factor 30 and restricted the fit to $\approx \SI{2.0}{\micro s}$ after the beginning of the pulse.
This procedure has a negligible impact on the determination of the parameters of interest,

We find that the slow component of the scintillation light, caused by the de-excitation of the triplet state of Ar excimers, has a characteristic time of \SI{789(13)}{ns}, deviating significantly from the \SI{1600}{ns} expected from literature. This decrease is caused by the contamination of LAr with traces of nitrogen or oxygen, which quenches the slow component of the scintillation light, causing an overall decrease of the scintillation yield.
The correction factor for the scintillation yield due to reduced LAr purity can be estimated as discussed in \cite{Acciarri:2009xj}
\begin{equation}
    f_\text{purity} = A_\text{s} + A_\text{t}\cdot\frac{\tau_\text{t}^\text{(exp)}}{\tau_\text{t}^\text{(pure)}}
    \label{eq:fpurity}
\end{equation}
where $A_\text{s}=0.23$ and $A_\text{t}=0.77$ indicates the relative population of singlet and triplet states for \ce{Ar^{*}_{2}} produced by {\small MIP}s, while $\tau_{t}^\text{(exp)}$ is the measured triplet state lifetime and $\tau_\text{t}^\text{(pure)} = \SI{1600}{ns}$ is the triplet state time constant for maximum light yield. 
The resulting $f_\text{purity}$ is \SI{0.61}{}, a sizeable suppression respect to high purity condition, but still sufficient for establishing the performance of the SoLAr anode plane design.

\subsection{Reconstruction of cosmic muons with charge data}
\label{sec:cosmics:charge}

One of the key tasks of the pixelated charge readout system on the anode plane is the reconstruction of particle tracks in the detector. 
Track reconstruction in SoLAr V2 builds upon well-established analysis tools, incorporated into a framework designed to first cluster charge hits and then fit tracks in three-dimensional space. The clustering of hits is achieved using the \textbf{D}ensity-\textbf{B}ased \textbf{S}patial \textbf{C}lustering of \textbf{A}pplications with \textbf{N}oise ({\tt DBSCAN})~\cite{ester1996density} algorithm, which groups spatially adjacent hits. Following this, the \textbf{Ran}dom \textbf{Sa}mple \textbf{C}onsensus ({\tt RANSAC})~\cite{bib:ransac} algorithm is applied to fit a track to the clustered hits. Both algorithms are inherently robust and, under ideal conditions, would be sufficient to reconstruct tracks on a uniform 3D grid of hits.

Inactive areas in the SoLAr V2 anode introduce additional challenges to the reconstruction process, requiring a more sophisticated approach. Figure\,\ref{fig:activeChanelMap} shows that the charge readout plane features some \textit{dead areas} due to malfunctioning LArPix chips, in the locations of SiPMs, and due to a few single defective channels. The resulting gaps in the charge collection cause  some clustering approaches to wrongly return fragmented tracks.

To address this effect, we employ a staged clustering procedure that not only identifies individual tracks but also bridges gaps where hits from a single track are missing due to the dead areas. This methodology, outlined in the following steps, is specifically designed to handle the challenges posed by the SoLAr V2 detector. The coordinate system used in the reconstruction has its origin at the center of the instrumented $4 \times 5$ sectors anode plane, with the $xy$ plane corresponding to the anode plane, and the $z$ axis pointing towards the cathode.

\begin{enumerate}
    \item \textit{Cluster hits on the $xy$ plane.}
    
    The \texttt{DBSCAN} algorithm is applied to the hits on the $xy$-plane using two key parameters: the ``minimum samples" threshold and $\epsilon_{xy}$. The minimum samples threshold ensures that the clusters contain at least two hits, while $\epsilon_{xy}$ defines the maximum spatial separation between hits. These parameters are adjustable and have been empirically optimised for this specific analysis. The optimal $\epsilon_{xy}$ is found to be \SI{8}{mm}, as it allows a separation of at most one pixel (\SI{4}{mm}) between hits. The primary function of this stage is noise suppression, removing isolated hits and those spaced too far apart to form a valid track segment. 

    \item \textit{Identify the $z$ axis intervals between clusters.}

    The clustering is repeated using the algorithm above but also in the drift direction. A new parameter $\epsilon_{z}$ is used to define the separation between hits and is also set to \SI{8}{mm}.
    
    \item \textit{Generate artificial space points within the dead areas on the $xy$ plane, constrained to the $z$ intervals identified in Step 2.}

    In this step, an interpolated cloud of space points is introduced to bridge the gap between clusters identified in Step 1. These points span from the highest hit in the lower cluster to the lowest hit in the upper cluster for each ordered pair of clusters, covering the dead areas in the $xy$-plane. Their spatial separation is set \SI{1}{mm} below $\epsilon_{xy}$ in the $xy$-plane and $\epsilon_z$ in the $z$-coordinate, ensuring that only physically meaningful clusters are formed in the next step. The charge is interpolated for these artificial clusters based on surrounding measured signals.

    \item \textit{Cluster the original and artificial data points in the $xy$-plane.}
    
    The \texttt{DBSCAN} algorithm is applied again, clustering both the original and artificial data points in the $xy$-plane. This step links the previously separate clusters into a continuous structure, using the artificial space points to fill the gaps.

    \item \textit{Refine clustering along the $z$-axis.}

    The clusters formed in Step 4 are then re-clustered along the $z$-axis using a maximum separation parameter $\epsilon_z$ of \SI{8}{mm}, similar to clustering in the XY-plane. 
    
    \item \textit{Remove artificial space points from the final clustered hits.}  
\end{enumerate}

Once the final clusters have been determined in Step 6, all artificial space points are removed, and the {\tt RANSAC} regression is applied. The charge of each hit is used as a weighting factor to prioritise regions with higher energy deposition. Unlike simple linear interpolation, {\tt RANSAC} selects the candidate track by iteratively fitting a regression model while rejecting outliers. Clusters that do not conform to the primary track are classified as outliers and can be reprocessed using the staged clustering method to identify secondary tracks if they exist. 

One limitation of the {\tt RANSAC} method is its non-deterministic nature. Given a sufficient number of iterations, bounded by the ``maximum trials" parameter, the algorithm should converge to a stable track solution. However, if the data are not well described by a linear model, the fit may fail. The goodness of a fit is quantified by the ``{\tt RANSAC} score", a normalised metric in the range $0 \leq \text{score} \leq 1$. Well-reconstructed tracks typically yield scores of $0.5 \leq \text{score} \leq 1.0$, while negative scores indicate that the model was unable to represent the data and should be discarded. 

The {\tt RANSAC} score does not directly measure the fit residuals. A high score may indicate strong alignment along one axis, while the distribution of charge data that forms the clusters remains scattered. In this study, the clustering procedure mitigates this issue to some extent, but in general, multiple metrics should be considered when selecting well-reconstructed tracks. Furthermore, an event may contain multiple tracks. To ensure signal quality, we consider only events in which a single track is reconstructed.

\subsubsection{Directionality and Track Length}
\label{sec:reco:directionality}

\begin{figure}[htpb]
    \centering
    \includegraphics[width=0.6\linewidth]{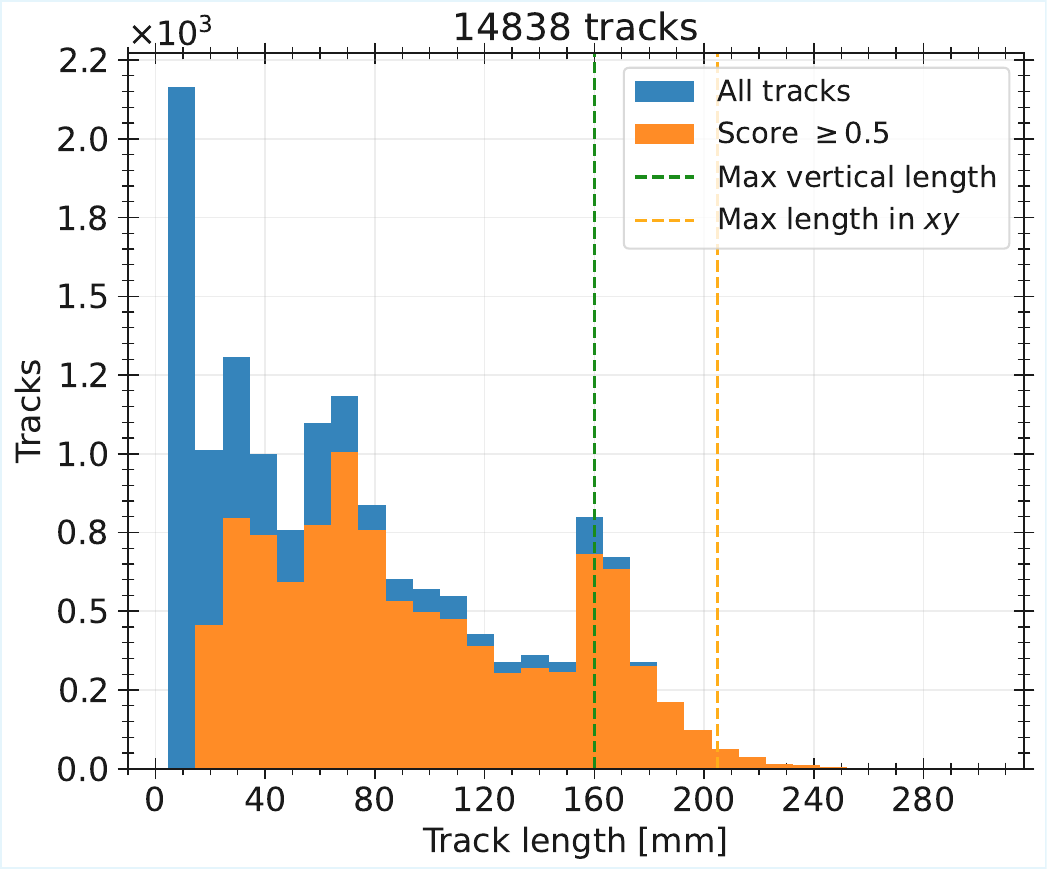}
    \caption{Distribution of track lengths for single-track events in data from cosmic-ray muon candidates. The two peaks correspond to the total height of the charge-sensitive pixel area in the $y$ axis, which is approximately 70 mm on the left side of the anode and 160 mm for the fully functional right side of the anode.}
    \label{fig:reco:track_length}
\end{figure}

The distribution of track lengths for single-track events is shown in Figure~\ref{fig:reco:track_length}. Many tracks shorter than \SI{16}{mm} are poorly reconstructed, often arising from noise or particles skimming the edges of the instrumented volume. We therefore exclude tracks below this length from the analysis. 
Above this threshold, two clear peaks appear, corresponding to well-reconstructed tracks with high {\tt RANSAC} scores. 

These two peaks arise from the maximum lengths achievable by vertical tracks (aligned with the \(y\)-axis), given the detector’s dead areas. Treating the instrumented region as a \(4\times 5\) array of unit cells, vertical tracks in the first column traverse \SI{64}{mm} of active anode. Tracks in the remaining three columns may intersect dead areas, but the reconstruction procedure described earlier bridges these gaps, restoring the full maximum track length of \SI{160}{mm}.

This hypothesis was tested by simulating a sample of cosmic muon tracks with \texttt{SOLAr-sim}, the SoLAr Monte Carlo package~\cite{SOLAr-sim}. The \texttt{SOLAr-sim} framework interfaces several event generators with Geant4~\cite{GEANT4:2002zbu, Allison:2006ve, Allison:2016lfl}, which propagates particle tracks through the user-defined detector geometry. It also computes the number of ionization electrons and scintillation photons produced in argon at each simulation step using the LArQL model~\cite{Marinho:2022xqk} and models the drift of the ionization cloud to the pixelated anode, providing an approximate description of the charge readout

\begin{figure}[htbp]
        \centering
        \includegraphics[width=0.7\linewidth]{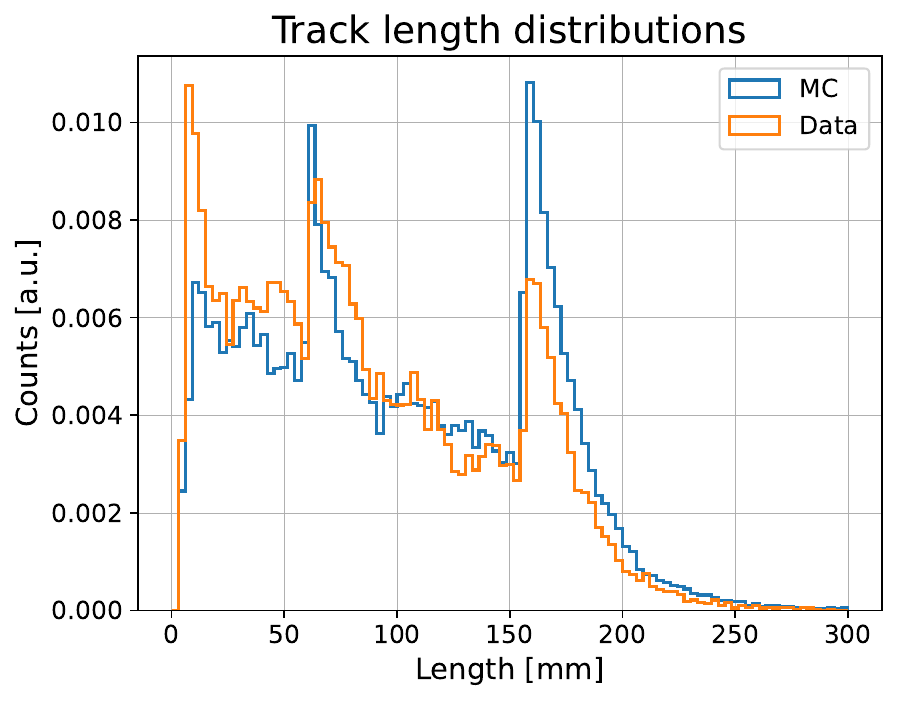}
    \caption{Comparison of reconstructed track length distributions for single-track events in data and Monte Carlo simulations, showing good agreement between observed and simulated reconstruction. The simulation includes inactive regions observed during data-taking.}
            \label{fig:reco:length_mc}
\end{figure}

For this study, cosmic-ray muons were simulated using the CRY event generator~\cite{CRY}, and the effect of dead areas on the anode was reproduced by disabling the corresponding channels in the simulation. The simulated events were analysed using the same procedure applied to the data. Figure~\ref{fig:reco:length_mc} compares the resulting track-length distributions from data and simulation. Although the relative amplitudes of the peaks differ, their positions agree well and support our interpretation.

\begin{figure}[htb]
    \centering
    \includegraphics[width=1.0\linewidth]{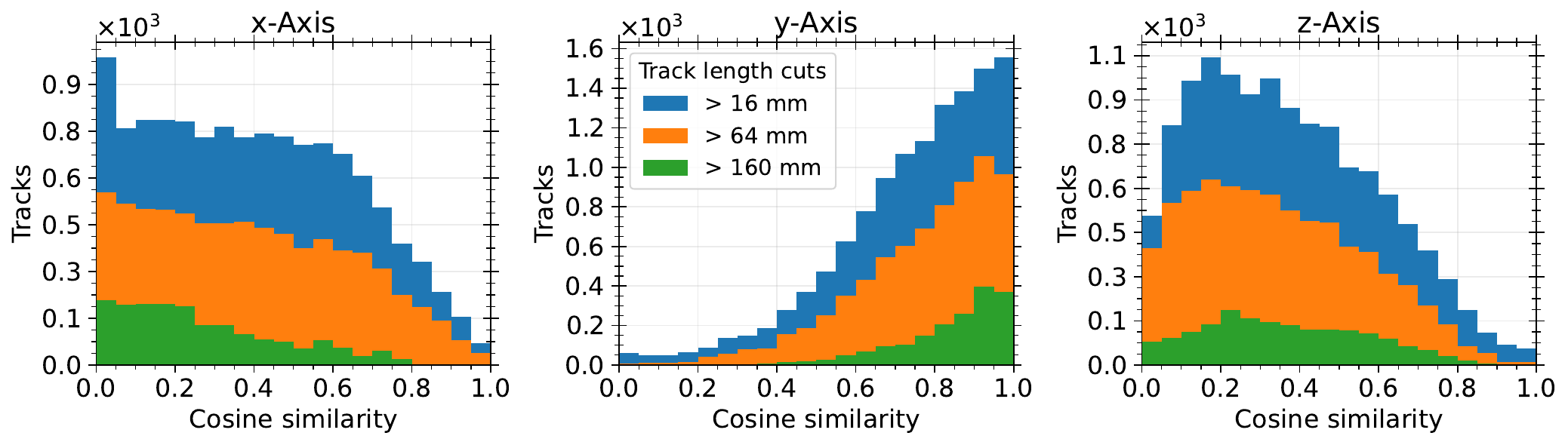}
    \caption{Angular distribution of tracks with length $>\SI{16}{mm}$, $>\SI{64}{mm}$, and $>\SI{160}{mm}$. Tracks are mostly vertical, consistent with cosmic ray trajectories.}
    \label{fig:reco:track_angle}
\end{figure}

This track-length pattern further supports the expectation that cosmic rays produce predominantly vertical tracks. Figure~\ref{fig:reco:track_angle} confirms this, showing a strong alignment of reconstructed tracks with the vertical axis. The observed spread arises from variations in track angle, short horizontal tracks, and tracks that do not fully traverse the detector.

\subsubsection{Charge Deposition per Unit Length and Electron Lifetime}
\label{sec:reco:lifetime}

Once a track is fitted, its charge deposition per unit length (\(\mathrm{d}Q/\mathrm{d}x\)) can be extracted for energy reconstruction. This is done by placing a sequence of equal-height cylinders along the track and summing the charge within each one. The cylinder height defines the target \(\mathrm{d}x\), while the collected charge in each cylinder gives \(\mathrm{d}Q\). The cylinder dimensions are set automatically using the track geometry and the clustering parameters. The cylinder height \(h\) is obtained by projecting the \texttt{DBSCAN} separation parameters \((\epsilon_{xy}, \epsilon_{xy}, \epsilon_z)\) onto the normalized track direction vector.

\begin{figure}[htbp]
\centering
\includegraphics[width=0.66\textwidth]{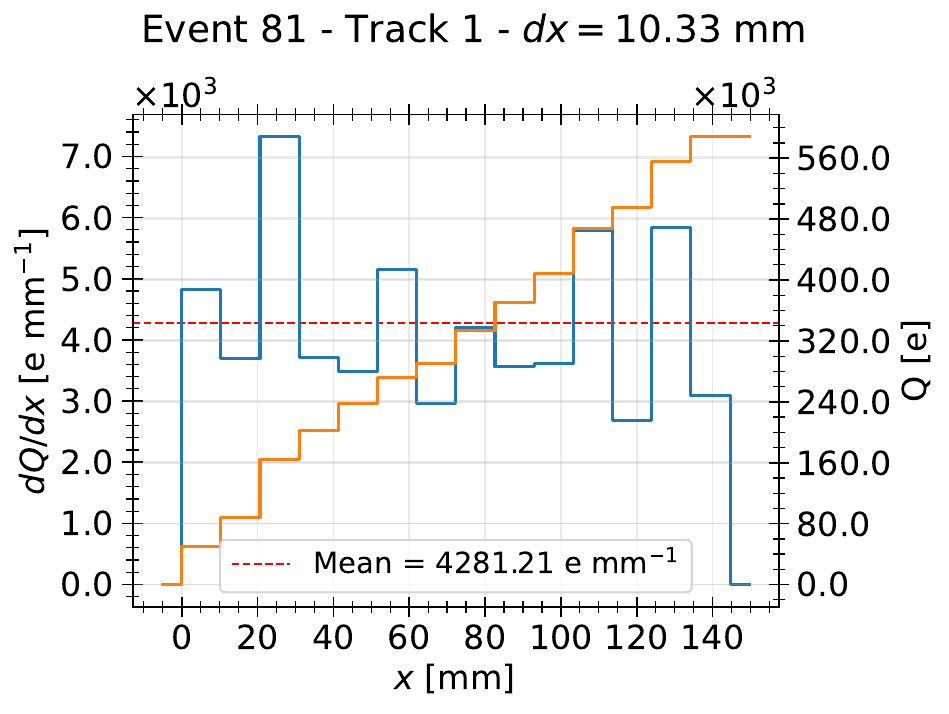}
\caption{Charge deposition per unit length ($\mathrm{d}Q/\mathrm{d}x$) profile for a single track with the cumulative total charge as a function of the track length (in orange). The mean $\mathrm{d}Q/\mathrm{d}x$) for this event is $4281$~e\,mm$^2$.}
\label{fig:ev81_dqdx} 
\end{figure}

The presence of dead areas in the detector introduces discontinuities in charge deposition, leading to artificially low $\mathrm{d}Q/\mathrm{d}x$ values in certain cylinders when $h$ is used directly as $\mathrm{d}x$. To correct for this, the effective $\mathrm{d}x$ is dynamically adjusted for each cylinder, taking into account the gaps in the sequence of hits. As a result, while $h$ determines the volume of the cylinder, the actual $\mathrm{d}x$ used in the $\mathrm{d}Q/\mathrm{d}x$ calculations reflects the cumulative displacement along the track line for each uninterrupted sequence of hits within the cylinder. The radius of the cylinder is determined from the fit residuals, using the root mean squared error of the fit:
\begin{equation}
    r = \sqrt{\sum^N_{i=1}\frac{\left|\vec{x}_i - \vec{p}_i\right|^2}{N-1}} \,
    \label{eq:reco:r}
\end{equation}  
where $\vec{x}_i$ represents the positions of individual hits, while $\vec{p}_i$ denotes their projected positions along the fitted track. The radius $r$ is subject to a lower bound of $|\vec{\epsilon}|/4$. Only hits within a cylinder contribute to the $\mathrm{d}Q/\mathrm{d}x$ calculations, while those at the interface between two cylinders are counted only once. For each track, we record the ratio of charge computed using this method to the total charge from all recorded hits. This ratio serves as an indicator of how much charge is incorporated into the $\mathrm{d}Q/\mathrm{d}x$ calculation relative to the total detected charge along the track.  
Figure~\ref{fig:ev81_dqdx} shows the resulting $\mathrm{d}Q/\mathrm{d}x$ profile for the track shown in Figure~\ref{fig:reco:ed}, obtained using this algorithm.

To estimate the total charge signal produced by a track, we interpolate the non-zero $\mathrm{d}Q/\mathrm{d}x$ values and extrapolate over the full track length to estimate charge in dead regions.
To accurately interpret the charge deposition profile, we must account for the capture of ionization electrons from argon impurities during the drift, which results in an overall attenuation of the charge signal.
This effect is parametrized by:
\begin{equation}
       \frac{\mathrm{d}Q}{\mathrm{d}x}\left(t_{\rm{drift}}\right)=\frac{\mathrm{d}Q_0}{\mathrm{d}x}\exp\left(\frac{-t_{\rm{drift}}}{\tau}\right),
    \label{eqn:lifetime}
\end{equation}
\noindent with $\tau$ representing the free electron lifetime in the drift volume and $t_{\rm{drift}}$ the drifting time.

\begin{figure}[hptb]
    \centering
    \includegraphics[width=0.6\linewidth]{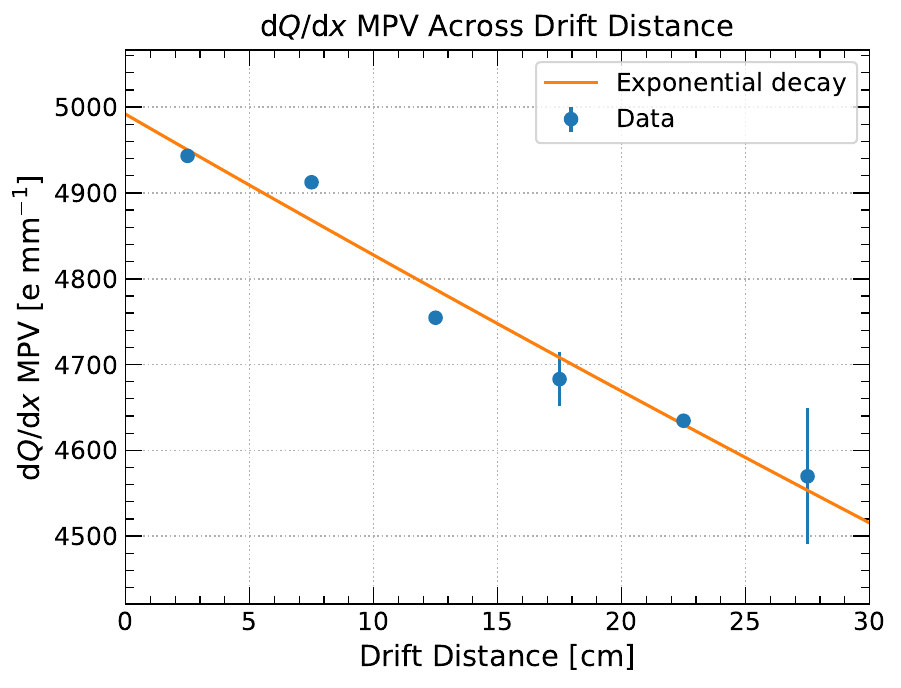}
    \caption{MPV values of the $\mathrm{d}Q/\mathrm{d}x$ as a function of drift distance. A fit to an exponential decay function obtained a drift electron lifetime of 1.87$\pm 0.18$ ms.}
    \label{fig:reco:lifetime}
\end{figure}

To estimate the electron lifetime, we group all reconstructed $\mathrm{d}Q/\mathrm{d}x$ values according to their distance from the readout plane, considering bins of \SI{50}{mm} along the drift length. For each dataset we fit the distribution of the $\mathrm{d}Q/\mathrm{d}x$ values with a Landau-Gauss distribution, which gives us the most probable value ({\small MPV}) for the charge deposition per unit length. 
Figure\,\ref{fig:reco:lifetime} shows the {\small MPV} obtained from the fit as a function of the drift distance.

\begin{figure}[hptb]
    \centering
    \includegraphics[width=0.6\textwidth]{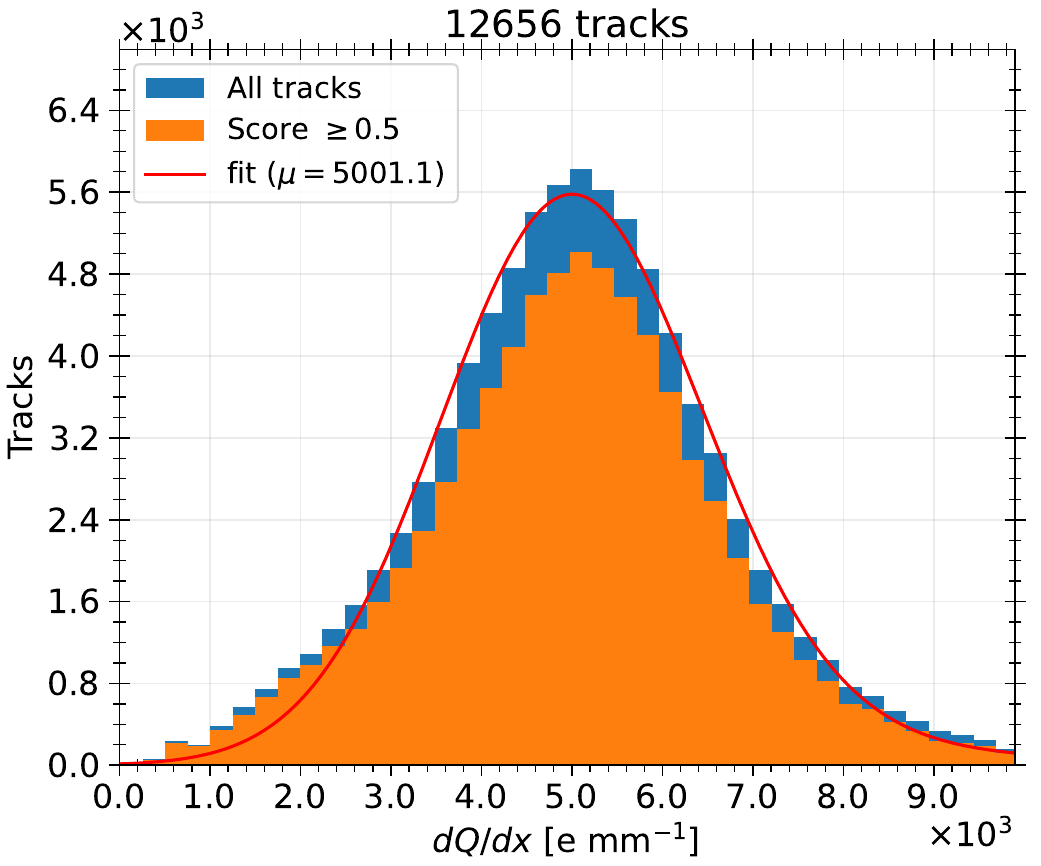}
    \caption{Lifetime-corrected $\mathrm{d}Q/\mathrm{d}x$ distribution for single-track events with a track length greater than \SI{16}{mm}.}
    \label{fig:reco:tracks_all}
\end{figure}

We determine an electron lifetime $\tau=\SI{1.87\pm 0.18}{ms}$, which corresponds to a maximum charge loss of $\approx 10\%$ for a \SI{30}{cm} drift length. 
The level of attenuation of the charge signals corresponds to an impurity concentration of 160 ppt \ce{O_2} equivalent. 
The DUNE collaboration observed a similar argon purity utilizing the same cryogenics infrastructure as the SoLAr V2 prototype~\cite{DUNE:2024fjn}.
Once the electron lifetime \(\tau\) is determined, we apply the corresponding correction to the raw \(\mathrm{d}Q/\mathrm{d}x\) values. The corrected \(\mathrm{d}Q/\mathrm{d}x\) distribution for all single-track events with lengths greater than \SI{16}{mm} is shown in Figure~\ref{fig:reco:tracks_all}.

A Landau--Gaussian fit to the calibrated calorimetric measurements yields a most probable value of \(\mathrm{MPV}=\SI{5.00}{ke.mm^{-1}}\) with a standard deviation of \(\sigma=\SI{1.37}{ke.mm^{-1}}\). A final gain correction of 6\% is applied to account for reduced charge-collection efficiency caused by SiPMs, which prevent some drifting electrons from reaching nearby pixels. This correction, derived from COMSOL electric-field simulations, increases the estimated \(\mathrm{d}Q/\mathrm{d}x\) to approximately \(\SI{5.30}{ke.mm^{-1}}\) for the LArPix pixels in the SoLAr V2 prototype, consistent with measurements from the LArPix collaboration~\cite{DUNE:2024fjn}.

\subsection{Matching Light Data to Charge Data from Reconstructed Muons}
\label{sec:reco:light}

Just as the pixelated anode provides a 3D coordinate system for charge hits, the SiPM array defines a 2D coordinate system for light ``hits.'' Because scintillation light in the TPC is emitted isotropically, there is no preferred direction, limiting the ability to reconstruct the full 3D position of the light source. However, the 2D pattern of light intensity recorded by the SiPMs remains proportional to the distance of the source from each SiPM on the anode plane.

Although SiPMs alone cannot directly determine the 3D coordinates of light hits, we can estimate the third coordinate by combining the $x$ and $y$ coordinates of the SiPMs with the $z$ coordinate of a nearby track reconstructed from charge hits. We select the five SiPMs with the strongest signals, with a minimum of three SiPMs having non-zero values. The $z$ coordinate for each SiPM is then estimated using one of the following methods:

\begin{enumerate}
    \item If charge hits are present within the same sector as the SiPM, the $z$ coordinate is taken as the weighted average of the charge hits $z$ coordinates, with weights proportional to the detected charge.
    \item If no charge hits are present, but a reconstructed track crosses the SiPM’s sector, the $z$ coordinate is taken as the height of the nearest point on the track to the SiPM.
    \item If neither charge hits nor a reconstructed track is available, the $z$ coordinate is estimated as the weighted average of the $z$-coordinates of all charge hits in the event.
\end{enumerate}

\subsection{Calorimetry of Muons with Charge and Light Signals}
\label{sec:reco:correlation}

One of the key goals of the SoLAr project consists in enhancing the energy resolution of LArTPCs by enabling combined calorimetry using the VUV-sensitive SiPMs integrated on the anode. 
We test the possibility of performing light-based calorimetry on a selected sample of cosmic-ray events,
focusing on those events where only one track, at least \SI{4}{cm} long, is reconstructed by the charge readout system. 
To ensure that the sample contained only well-reconstructed tracks, we also require a ``\texttt{RANSAC} score'' larger than 0.5. 

Different from the measured charge signal, which is produced only in the region covered by the instrumented part of the anode plane, detected photons may have originated anywhere along the track crossing the TPC. 
For this reason, we propagate each selected track within the TPC volume until reaching its entry and exit point ($\vec{x}_\text{in}$ and $\vec{x}_\text{out}$ respectively).

\begin{figure}
    \centering
    \includegraphics[width=0.6\linewidth]{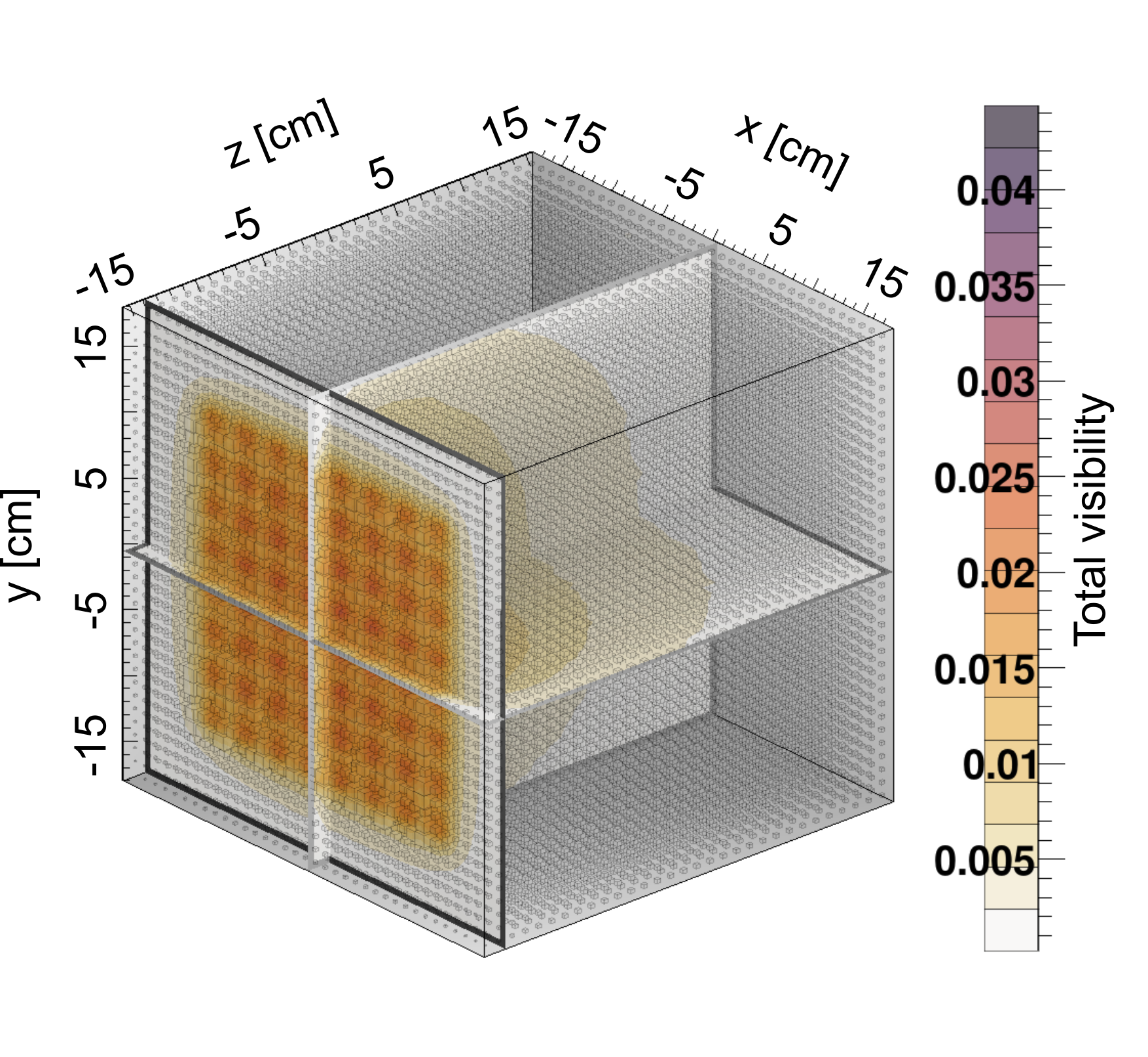}
    \caption{%
        SoLAr anode total visibility, defined as the sum of the each SiPM contributions $\sum_i^{{N}_\text{SiPM}} \Omega_i(\vec{x})$.
        The cross section views help accentuate the visibility throughout the inside of the TPC volume.
    }
    \label{fig:vmap}
\end{figure}

Any calorimetric estimate using the observed light signal as an input requires a light propagation model to take into account the
optical properties of the medium and the particular topology of the event under consideration. 
Given the small size of the SoLAr V2 TPC, we build a LookUp Table (LUT) using the SoLAr Monte Carlo simulation to propagate LAr scintillation photons originating from a 3D grid of points with a spacing of \SI{1}{cm}. 
The simulation includes photon absorption and Rayleigh scattering, while it does not include reflection on the cathode or on the field-shaping rings of the TPC. 
For each simulated vertex $\vec{x}$, we fill the LUT with the \emph{visibility} of each SiPM $\Omega(\vec{x})$, i.e., the fraction of photons collected by each sensor. A 3D representation of the SoLAr total anode visibility, defined as the sum of the individual contributions from all SiPMs,  is shown in Figure\,\ref{fig:vmap}. 

We compute the expected signal $q_i$ on the $i$-th SiPM by propagating the reconstructed track in the TPC volume and integrating the light-emission along the track
\begin{equation}
    q_i = \sum_{k=1}^{N} 
    \left.\frac{\textrm{d}E}{\textrm{d}x}\right|_\text{MIP} \cdot \delta x
    \cdot \left(\text{LY}(\mathcal{E})\cdot f_\text{purity}\right)
    \cdot \text{PDE} 
    \cdot \Omega_i\left(\vec{x}_\text{in} + k\,\delta x\,\hat{u}_\text{track}\right) \;,
    \label{eq:qsipm0}
\end{equation}
where $\textrm{d}E/\textrm{d}x|_\text{MIP}$ is the characteristic energy loss of a MIP per unit length, $\delta x$ is the integration step, $\text{LY}(\mathcal{E})$ indicates the expected light yield, \textit{i.e.}, the expected number of photons emitted per unit of deposited energy under the electric field $\mathcal{E}$, $f_\text{purity}$ is the purity correction factor introduced in \eqref{eq:fpurity}, PDE denotes the average SiPM photodetection efficiency, and $\hat{u}_\text{track}$ indicates the reconstructed track direction unit vector. As the sum runs across the entire track length $\ell_\text{track}^\text{TPC} = |\vec{x}_\text{in}-\vec{x}_\text{out}|$, the summation index becomes $N = \ell_\text{track}^\text{TPC} / \delta x$. 

\begin{figure}
    \centering
    \begin{subfigure}[b]{0.48\linewidth}
     \centering
         \includegraphics[height=6.2cm]{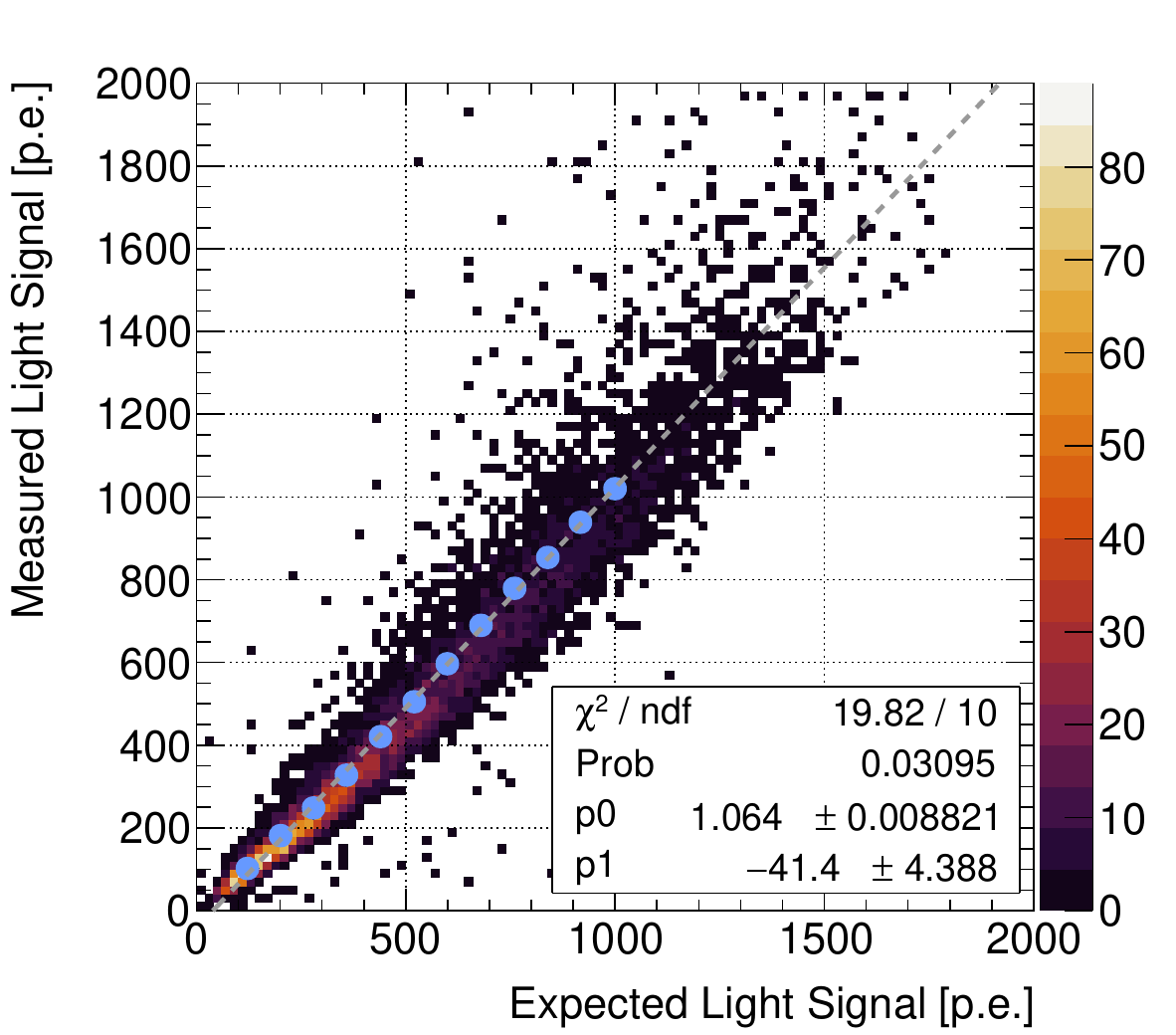}
         \caption{}
         \label{fig:meas_vs_exp_light}
     \end{subfigure}
     \hfill
     \begin{subfigure}[b]{0.48\linewidth}
     \centering
         \includegraphics[height=6.2cm]{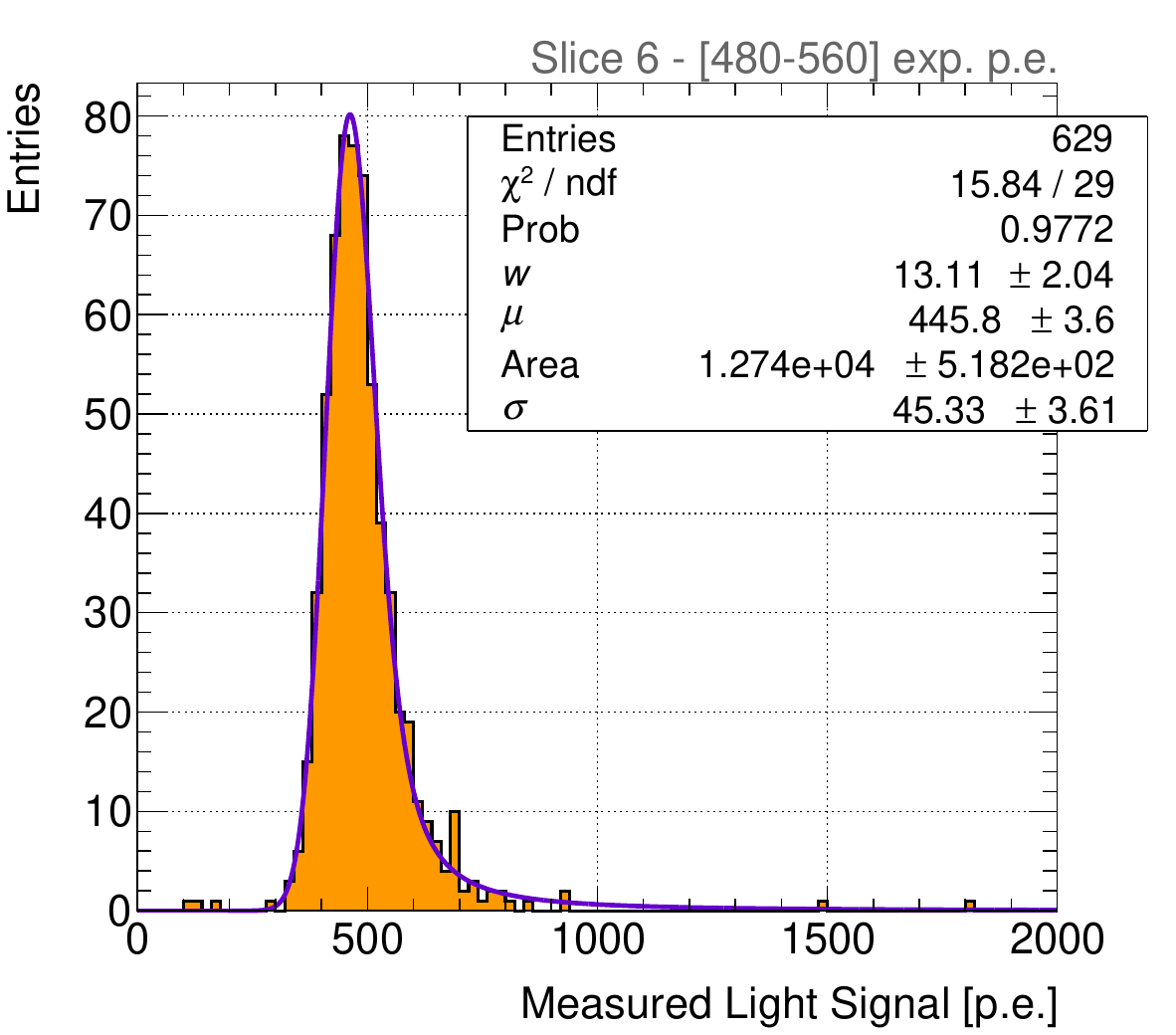}
         \caption{}
         \label{fig:slice_fit}
     \end{subfigure}
    \caption{%
        (a) Correlation between the measured and expected light signal in SoLAr V2. 
        (b) Example fit of the measured light signal distribution obtained selecting 
        events with $L_\text{exp}\in[480, 560]\,\text{p.e.}$.
    }
    \label{fig:placeholder}
\end{figure}

Defining an average visibility $\overline{\Omega}_i(\vec{x}_\text{in}, \hat{u}_\text{track}) = \frac{1}{N}\sum_{k=1}^{N} \Omega_i\left(\vec{x}_\text{in} + k\,\delta x\,\hat{u}_\text{track}\right)$, we then use Eq.\,\eqref{eq:qsipm0} to estimate the expected light signal $L_\text{exp}$ on the entire SoLAr anode as
\begin{equation}
    L_\text{exp} =
    \left.\frac{\textrm{d}E}{\textrm{d}x}\right|_\text{MIP} \cdot \ell_\text{track}^\text{TPC}
    \cdot \left(\text{LY}(\mathcal{E})\cdot f_\text{purity}\right)
    \cdot \text{PDE} 
    \cdot \sum_{i=1}^{N_\text{SiPM}}\overline{\Omega}_i\left(\vec{x}_\text{in}, \hat{u}_\text{track}\right) 
    \label{eq:Lexp} 
\end{equation}
with $N_\text{SiPM}$ indicating the number of active SiPMs in the event. Figure\,\ref{fig:meas_vs_exp_light} shows the correlation between the measured light signal $L_\text{data}$ and $L_\text{exp}$, computed assuming \SI{2.1}{MeV\,cm^{-1}} as  $\textrm{d}E/\textrm{d}x|_\text{MIP}$ mean value \cite{lar_webpage}, an expected light yield of \SI{25000}{ph\,MeV^{-1}} \cite{Marinho:2022xqk}, and a PDE of \SI{12.7}{\%} as measured in Ref.~\cite{Alvarez-Garrote:2024byb}.

To determine the calibration factor between the expected and measured light signal, the full $L_\text{data}$ sample is divided into 12 sub-sets selected in 80\,p.e.-wide intervals of $L_\text{exp}$. 
For each sub-sample, the distribution of $L_\text{data}$ is fitted with a Landau distribution convolved with a Gaussian to account for the the limited resolution of the detector. One example of such analysis is shown in Figure\,\ref{fig:slice_fit}. 

From every fit, we compute the mean value of the Landau-Gaussian distribution numerically. The corresponding  values of $L_\text{exp}$ are obtained from the mean of the $L_\text{exp}$ distribution within each interval. 
These results are shown as light-blue markers in Fig.\,\ref{fig:meas_vs_exp_light}. 
The uncertainty associated to each $L_\text{exp}$ value is the standard deviation of the mean,
while uncertainties on $L_\text{data}$ are evaluated drawing 100 sets of correlated parameters using the fit covariance matrix and repeating the numerical evaluation of the mean value for each extraction. The standard deviation of the resulting distribution is then the used to quantify the uncertainty for the $L_\text{data}$ value.

The trend of $L_\text{data}$ as a function of $L_\text{exp}$ is consistent with a linear model, and the slope value close to 1 indicates that our approach describes the overall light response with sufficient accuracy.
However, the relatively large intercept suggests that the model does not capture entirely the behaviour of the detector. In particular, the agreement can be improved by including a more detailed description of the optical properties of the detector surfaces, an angular dependence of the SiPM PDE, and an improved calibration that accounts for the SiPM afterpulses.

Crossing the small TPC volume as MIPs, cosmic muons are not an ideal sample to study the detector energy resolution. 
Nevertheless, the good agreement between the observed light signal and the light propagation model allows us to estimate, analogously to $\textrm{d}Q/\textrm{d}x$, the characteristic light emission per unit length $\textrm{d}L/\textrm{d}x$ defined as 
\begin{equation}
    \frac{\textrm{d}L}{\textrm{d}x} = \left.\frac{\textrm{d}E}{\textrm{d}x}\right|_\text{MIP}
    \cdot \left(\text{LY}\cdot f_\text{purity}\right) = 
    \frac{L_\text{data}}{%
        \ell_\text{track}^\text{TPC}
        \cdot \text{PDE} 
        \cdot \sum_{i=1}^{N_\text{SiPM}}\overline{\Omega}_i\left(\vec{x}_\text{in}, \hat{u}_\text{track}\right) \; .
        }
    \label{eq:dLdx}
\end{equation}

\begin{figure}
    \centering
    \includegraphics[width=0.85\linewidth]{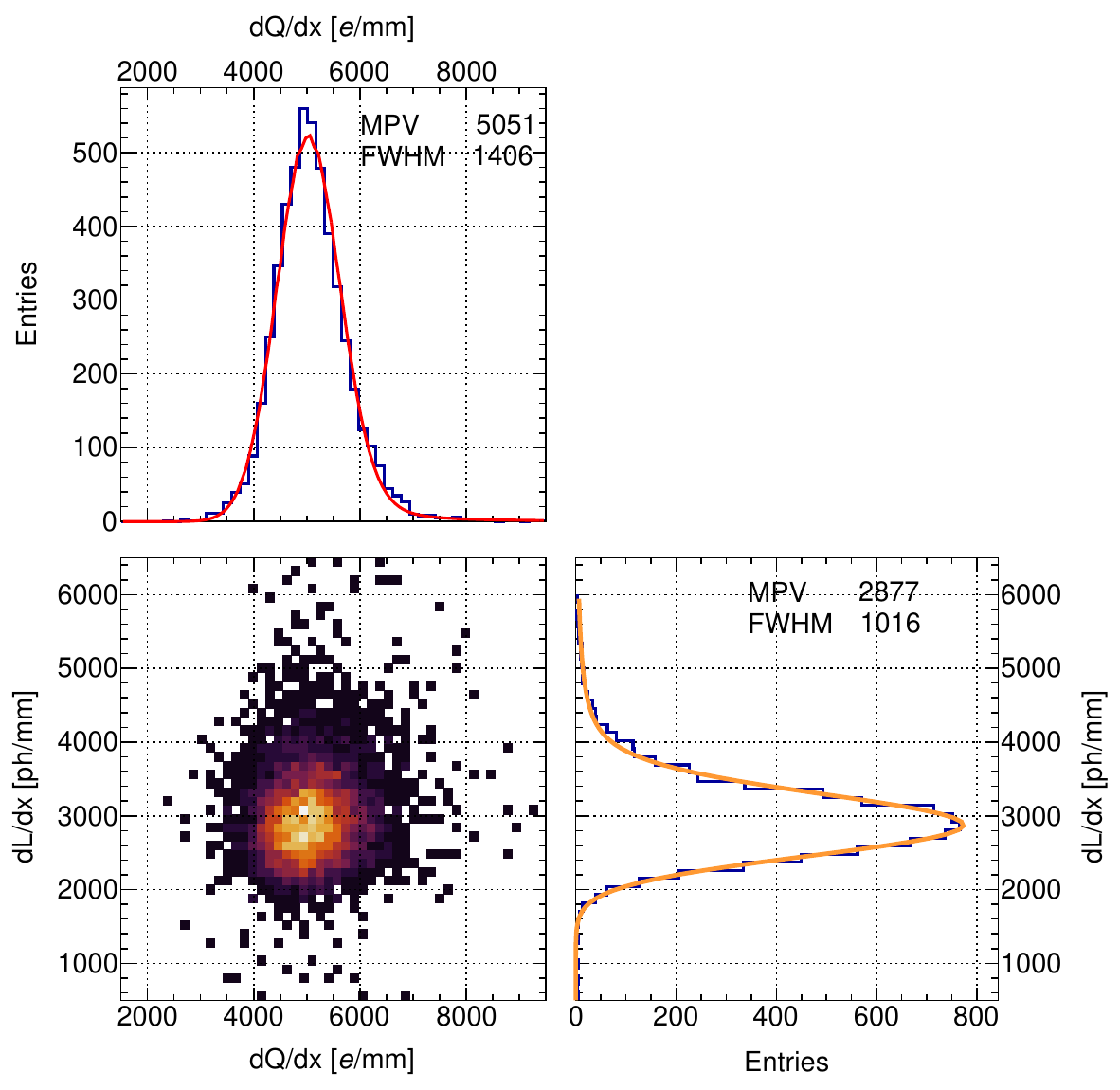}
    \caption{%
        Joint and marginalized distributions of $\textrm{d}Q/\textrm{d}x$ and $\textrm{d}L/\textrm{d}x$ for a selected sample of cosmic muons tracks crossing the TPC.}
    \label{fig:dQdLcorr}
\end{figure}

In Figure\,\ref{fig:dQdLcorr} we show the joint distribution of $\textrm{d}L/\textrm{d}x$ and $\textrm{d}Q/\textrm{d}x$ along with their marginalized distributions, which have been fitted with a Landau distribution convolved with a Gaussian function. Consistent results are shown for cosmic-ray muons for both light and charge. 

The two distributions feature a similar relative width, with $\text{FWHM}/\text{MPV}$ being $28\%$ and $35\%$ for $\textrm{d}Q/\textrm{d}x$ and $\textrm{d}L/\textrm{d}x$, respectively. The intrinsic $\text{d}E/\text{d}x$ spread, determined with a Monte Carlo simulation, is below $10\%$ for both the charge and light sample, suggesting comparable performances for charge and light-based calorimetry.

A more accurate assessment of the energy resolution of the SoLAr light readout can be obtained by using contained events of known energy.
During the same data taking period, we acquired a dedicated data set with a \ce{^{60}Co} radioactive source placed outside of the cryostat to probe the MeV-scale performance of the SoLAr detector concept.
The results of this run will be presented in a future publication, which will include a more detailed study of the energy resolution gain achievable by combining charge and light signals.

\section{Conclusion}
The paper describes the second test of the SoLAr readout concept, an innovative pixelated anode tile integrating charge and light sensitive components for the readout of LArTPCs. 
The SoLAr V2 tile houses 64 VUV-sensitive SiPMs and up to 64 LArPix ASICs for the readout of the charge collection pads. An area of $\SI{12.8}{}\times\SI{16.0}{cm^2}$ was instrumented for charge readout using 20 LArPix ASICs, marking a factor 5 increase with respect to the prototype used in the first SoLAr demonstration run~\cite{SoLAr:2024fwt}.

In July 2023 the anode was deployed into a $30\times30\times30\,\si{cm^3}$ LArTPC at the University of Bern, collecting a sample of tens of thousands of cosmic-muon tracks. 
We describe the calibration procedures for the charge and light readout system, focusing on the characterization of the SiPMs' response performed using dedicated LED calibration runs. 

We developed a track-reconstruction procedure and presented performance metrics for both track topology and charge-based energy loss. We also studied the impact of a highly segmented light-readout system on the anode, constructing a simplified model to estimate the deposited energy from LAr scintillation light using the track topology obtained from the charge readout. Applying this method to cosmic-ray muon data demonstrates that the light collected by the SoLAr anode can provide a valuable handle for improving the energy resolution of LArTPCs.

Future studies will leverage these results to develop analysis procedures for low-energy data samples collected during the same data taking period exposing the TPC to a radioactive source. This dataset will allow us to test the reconstruction performance of the SoLAr readout concept at an energy scale comparable to the one solar neutrino events in LAr. 
In particular we will study the impact of the combined calorimetry approach using both charge and light signal at the MeV scale, a crucial test of the SoLAr strategy for enhancing the calorimetric performance and improved background rejection of LArTPCs at low energy.

\section*{Acknowledgements}

This document was prepared by the SoLAr collaboration. The project has received funding from the European Union’s Horizon 2020 Research and Innovation programme under GA no 101004761, by the Italian Ministry for Research and University (MUR) under Grant Progetto Dipartimenti di Eccellenza 2023-2027 and the PRIN 2022 program (grant number 2022STFALX ``Beyond Liquid Argon (BeLAr)''), by the University of Milano-Bicocca (grants FAQC 2022 and 2024), by CERN DRD2 through INFN, by the Swiss National Science Foundation, by MICIU/AEI/10.13039/501100011033 and FEDER/UE (PID2023-147949NB-C51 grant). It has also been supported by several of the collaborating institutions: CIEMAT; Lawrence Berkeley National Laboratory; University of Bern; University of Manchester; University of Milano-Bicocca and INFN Sezione di Milano-Bicocca.

\bibliographystyle{JHEP}
\bibliography{biblio.bib}

\end{document}